\documentclass[lettersize,journal]{IEEEtran}
\usepackage{amsmath,amsfonts,amssymb}
\usepackage{algorithmic}
\usepackage{algorithm}
\usepackage{array}
\usepackage[caption=false,font=normalsize,labelfont=scriptsize,textfont=scriptsize]{subfig}
\usepackage{textcomp}
\usepackage{booktabs}
\usepackage{enumitem}
\usepackage{stfloats}
\usepackage{url}
\usepackage{verbatim}
\usepackage{graphicx}
\usepackage{cite}
\usepackage{color}
\begin{document}

\title{Stacked Intelligent Metasurfaces for Efficient Holographic MIMO Communications in 6G}
\author{Jiancheng~An,~\IEEEmembership{Member,~IEEE,}
Chao~Xu,~\IEEEmembership{Senior~Member,~IEEE,}
Derrick~Wing~Kwan~Ng,~\IEEEmembership{Fellow,~IEEE,}
George~C.~Alexandropoulos,~\IEEEmembership{Senior~Member,~IEEE,}
Chongwen~Huang,~\IEEEmembership{Member,~IEEE,}
Chau~Yuen,~\IEEEmembership{Fellow,~IEEE,}
and~Lajos~Hanzo,~\IEEEmembership{Life~Fellow,~IEEE}
\thanks{This research is supported by the Ministry of Education, Singapore, under its MOE Tier 2 (Award number MOE-T2EP50220-0019). Any opinions, findings and conclusions or recommendations expressed in this material are those of the author(s) and do not reflect the views of the Ministry of Education, Singapore. This research is supported by A*STAR under its RIE2020 Advanced Manufacturing and Engineering (AME) Industry Alignment Fund – Pre Positioning (IAF-PP) (Grant No. A19D6a0053). Any opinions, findings and conclusions or recommendations expressed in this material are those of the author(s) and do not reflect the views of A*STAR. D. W. K. Ng is supported by the Australian Research Council's Discovery Project (DP210102169, DP230100603). The work of Prof. Alexandropoulos has been supported by the SNS JU TERRAMETA project under EU’s Horizon Europe research and innovation programme under Grant Agreement No 101097101. The work of Prof. Huang was supported by the China National Key R\&D Program under Grant 2021YFA1000500, National Natural Science Foundation of China under Grant 62101492, Zhejiang Provincial Natural Science Foundation of China under Grant LR22F010002, Zhejiang University Global Partnership Fund, Zhejiang University Education Foundation Qizhen Scholar Foundation, and Fundamental Research Funds for the Central Universities under Grant 2021FZZX001-21. L. Hanzo would like to acknowledge the financial support of the Engineering and Physical Sciences Research Council projects EP/W016605/1 and EP/X01228X/1 as well as of the European Research Council's Advanced Fellow Grant QuantCom (Grant No. 789028).}
\thanks{J. An is with the Engineering Product Development (EPD) Pillar, Singapore University of Technology and Design (SUTD), Singapore 487372 (e-mail: jiancheng\_an@sutd.edu.sg).}
\thanks{C. Xu, and L. Hanzo are with the School of Electronics and Computer Science, University of Southampton, Southampton SO17 1BJ, U.K. (e-mail: cx1g08@soton.ac.uk; lh@soton.ac.uk).}
\thanks{D. W. K. Ng is with the School of Electrical Engineering and Telecommunications, University of New South Wales (UNSW), Sydney, NSW 2052, Australia (e-mail: w.k.ng@unsw.edu.au).}
\thanks{G. C. Alexandropoulos is with the Department of Informatics and Telecommunications, National and Kapodistrian University of Athens, 15784 Athens, Greece (e-mail: alexandg@di.uoa.gr).}
\thanks{C. Huang is with College of Information Science and Electronic Engineering, Zhejiang University, Hangzhou 310027, China, with the State Key Laboratory of Integrated Service Networks, Xidian University, Xi’an 710071, China, and with Zhejiang-Singapore Innovation and AI Joint Research Lab and Zhejiang Provincial Key Laboratory of Info. Proc., Commun. \& Netw. (IPCAN), Hangzhou 310027, China (e-mail: chongwenhuang@zju.edu.cn).}
\thanks{C. Yuen is with the School of Electrical and Electronics Engineering, Nanyang Technological University, Singapore 639798 (e-mail: chau.yuen@ntu.edu.sg).}}
\maketitle
\begin{abstract}
The revolutionary technology of \emph{Stacked Intelligent Metasurfaces (SIM)} has been recently shown to be capable of carrying out advanced signal processing directly in the native electromagnetic (EM) wave domain. An SIM is fabricated by a sophisticated amalgam of multiple stacked metasurface layers, which may outperform its single-layer metasurface counterparts, such as reconfigurable intelligent surfaces (RISd) and metasurface lenses. We harness this new SIM concept for implementing efficient holographic multiple-input multiple-output (HMIMO) communications that dot require excessive radio-frequency (RF) chains, which constitutes a substantial benefit compared to existing implementations. We first present an HMIMO communication system based on a pair of SIMs at the transmitter (TX) and receiver (RX), respectively. In sharp contrast to the conventional MIMO designs, the considered SIMs are capable of automatically accomplishing transmit precoding and receiver combining, as the EM waves propagate through them. As such, each information data stream can be directly radiated and recovered from the corresponding transmit and receive ports. Secondly, we formulate the problem of minimizing the error between the actual end-to-end SIMs'parametrized channel matrix and the target diagonal one, with the latter representing a flawless interference-free system of parallel subchannels. This is achieved by jointly optimizing the phase shifts associated with all the metasurface layers of both the TX-SIM and RX-SIM. We then design a gradient descent algorithm to solve the resultant non-convex problem. Furthermore, we theoretically analyze the HMIMO channel capacity bound and provide some useful fundamental insights. Extensive simulation results are provided for characterizing our SIM-based HMIMO system, quantifying its substantial performance benefits. Indicatively, it is demonstrated that a $150$\% capacity improvement is feasible when compared with MIMO and RIS-aided communication systems.
\end{abstract}


\section{Introduction}
\IEEEPARstart{W}{ith} the completion of the 3GPP Release 17, it is high time for both industry and academia to begin conceptualizing the sixth-generation (6G) mobile networks \cite{3gpp_rel17}. Wireless network evolution has been primarily motivated by the pursuit of higher data rates and wider device connectivity. While this demand will continue to increase, the explosive proliferation of the Internet-of-Everything (IoE), ranging from extended reality to interconnected autonomous systems, is driving a fundamental paradigm shift \cite{samsung_6G}. It is envisaged that by 2030, the number of connected devices will reach $500$ billion, according to Cisco's annual report \cite{cisco}. To support these heterogeneous IoE services imposing extreme performance requirements, 6G wireless networks are expected to integrate communication, sensing, computing, and control capabilities, while drastically improving data rates, latency, and connectivity \cite{NE_2020_Dang_what}. As such, 6G will undergo a revolutionary transformation by flexibly orchestrating both physical and virtual resources to support the envisioned heterogeneous IoE scenarios and by harnessing sophisticated disruptive techniques, including artificial intelligence (AI) \cite{CM_2019_Letaief_The} spanning all network layers \cite{Alexandropoulos2022Pervasive}, satellite communications \cite{CST_2021_Kodheli_Satellite}, programmable reflective~\cite{WC_2022_An_Codebook} and computing metasurfaces \cite{Computing_metasurfaces2023}, and integrated sensing and communications devices and techniques \cite{arXiv_2023_An_Fundamental, IoTJ_2023_Xu_OTFS}, to name a few.
\subsection{Emerging Metasurface-Based Technologies}
We commence by reviewing a pair of 6G enabling techniques, namely, reconfigurable intelligent surfaces (RISs) and holographic multiple-input multiple-output (HMIMO) communications.
\subsubsection{Reconfigurable Intelligent Surfaces (RISs)}
The emerging metasurface technology was shown to be able to shape smart reconfigurable environments \cite{Strinati2021Reconfigurable,CM_2020_Wu_Towards, WC_2022_An_Codebook, WC_2021_Yu_Smart, CM_2022_Zhang_Intelligent, ICN_2022_Jian_Reconfigurable}. Specifically, a programmable metasurface is composed of a large number of low-cost passive reflecting elements, which are capable of manipulating the electromagnetic (EM) behavior of radio waves~\cite{WavePropTCCN}, allowing for proactive customization of the wireless propagation environment \cite{TCOM_2022_An_Low, CM_2021_Alexandropoulos_Reconfigurable}. By employing an RIS, \cite{TWC_2019_Huang_Reconfigurable} designed a scheme that substantially boosted the energy efficiency of the downlink in multiuser multiple-input single-output (MISO) communication systems, as compared to conventional relaying solutions. Following this pioneering work, a great deal of research has sprouted up by investigating the effects of realistic hardware imperfections \cite{TCOM_2021_Wu_Intelligent} and RIS element responses~\cite{TCCN_2022_Xu_Time}, optimizing the RIS phase shifts for wideband operation~\cite{TVT_2023_Xu_Channel} and based on the statistical channel state information (CSI) \cite{WCL_2022_Xu_Deep, TVT_2019_Han_Large}, and enhancing the quality-of-service (QoS) \cite{TVT_2022_Xu_Reconfigurable, TWC_2020_Pan_Multicell, WCL_2022_An_Scalable} in RIS-assisted communication systems. However, the encouraging performance benefits of these RIS solutions rely on the availability of accurate CSI, which generally requires excessive pilot overhead for acquisition \cite{ICN_2022_Jian_Reconfigurable}. To address this issue, the authors in \cite{Decoupling_Ill_2022,TCOM_2022_An_Low} devised codebook-based frameworks for channel estimation in RIS-assisted MIMO systems. According to those frameworks, the estimation of the RIS-parametrized channel and the design of transmit beamforming were handled via conventional protocols, while simplifying the reflection coefficient optimization by selecting the best entry from otpimized RIS reflection beam codebooks \cite{WCL_2021_An_The, WCL_2023_Jia_Environment}. It has been shown that the designed codebook-based solutions are appealingly scalable exhibiting strong robustness against hardware imperfections \cite{WCL_2021_An_The, TCOM_2022_An_Low}.
\subsubsection{Holographic MIMO (HMIMO)}
Over the past decade, the massive MIMO technique has become one of the most crucial enablers for increasing wireless capacity \cite{CM_2014_Larsson_Massive}. Explicitly, massive MIMO has the potential of focusing energy into a smaller spatial region, thus attaining huge improvements in both spectral and energy efficiencies \cite{TWC_2010_Marzetta_Noncooperative}. It also provides other benefits, such as it can be realized with inexpensive low-power components, exhibit reduced latency, lead to simplified protocol designs, and be robustness against jamming \cite{CM_2014_Larsson_Massive}. As the 6G research is ramping up, a natural question arises -- \emph{what will the next generation MIMO be like?} Recently, the innovative concept of HMIMO has emerged \cite{WC_2020_Huang_Holographic, TSP_2018_Hu_Beyond, WC_2021_Shlezinger_Dynamic,HMIMO_Survey_2023}. Specifically, by employing a large intelligent surface (LIS) constructed of an electromagnetically active material that integrates massive numbers of radiating and sensing elements\cite{TWC_2023_Xu_Antenna}, impressive improvements are expected \cite{TCOM_2021_Wan_Terahertz}. Furthermore, \cite{JSAC_2020_Dardari_Communicating} demonstrated that LIS-aided solutions are capable of improving the spatial multiplexing gain even in strong line-of-sight (LoS) propagation conditions. To characterize the fundamental capacity limits of HMIMO communications, the authors in \cite{JSAC_2020_Pizzo_Spatially} established a mathematically tractable channel model by considering the small-scale fading in the far-field as a spatially correlated random Gaussian field, while being consistent with the scalar Helmholtz equation. The same authors then developed a Fourier plane-wave series-based expansion of the HMIMO channel response for arbitrary scattering environments \cite{TWC_2022_Pizzo_Fourier}. Following this channel model, a channel estimation scheme leveraging the specific array geometry for identifying a low-dimensional common subspace for arbitrary spatial correlation matrices was designed in \cite{WCL_2022_Demir_Channel}. Moreover, the authors in \cite{CL_2022_Hu_Holographic} studied the family of discrete amplitude-controlled holographic beamformers with the specific objective of satisfying a given sum-rate requirement, while multi-user HMIMO systems were investigated in \cite{JSAC_2022_Deng_HDMA, JSTSP_2022_Wei_Multi}.
\subsection{Motivation}
It has been recently proposed to cascade multiple metasurfaces\footnote{A range of other terminologies having a similar multilayer structure were also used in different research communities, such as programmable AI machine \cite{NE_2022_Liu_A}, stacked metasurface slab \cite{ICAMNWP_2018_Chamanara_Stacked}, 3D integrated metasurface device \cite{LSA_2019_Hu_3D}, cascaded metasurfaces \cite{APL_2013_Pfeiffer_Cascaded}, and multilayer metasurfaces \cite{NL_2018_Zhou_Multilayer}.} to realize \emph{stacked intelligent metasurfaces (SIMs)}, which have the capability to implement signal processing in the EM wave regime \cite{ICC_2023_An_Stacked, ICAMNWP_2018_Chamanara_Stacked, NE_2022_Liu_A}. In this paper, we propose the integration of SIMs with the transceivers to support HMIMO communications. Before proceeding, we first elaborate on our motivation by answering the question -- \emph{why do we need SIM?} -- from the following three perspectives:
\begin{enumerate}
 \item The existing research on HMIMO communications is still in its infancy and lacks practical implementations~\cite{HMIMO_Survey_2023}. Since integrating an abundance of expensive active elements at the transceiver is an impractical option, recent research efforts focus on implementing HMIMO communications by employing programmable metasurfaces \cite{WC_2020_Huang_Holographic, TSP_2018_Hu_Beyond,WC_2021_Shlezinger_Dynamic}. However, their performance remains limited by practical hardware constraints, such as the tunable amplitude/phase associated with each meta-atom of a single-layer metasurface. As a remedy, a multilayer metasurface architecture might be beneficial for improving both the spatial-domain gain and the design degrees of freedom, thus, flexibly forming diverse radio frequency (RF) waveforms compared to its single-layer counterparts.
 \item Although integrating RISs into existing wireless networks has been numerically shown to improve both the spectral and energy efficiencies in various scenarios \cite{TGCN_2022_An_Joint, TWC_2019_Huang_Reconfigurable, TCOM_2021_Wu_Intelligent}, there are still stumbling blocks in the way of practical deployments of RISs. On one hand, the two-hop multiplicative path loss coefficient severely impacts the resultant performance \cite{TWC_2021_Tang_Wireless}. Indeed, several works investigate the RIS placement to reduce pathloss \cite{TCOM_2021_Wu_Intelligent} or optimize a certain performance objective \cite{moustakas_2023}. As a further impediment, the widespread RIS deployment significantly increases the burden of media access control (MAC) optimization, including both resource allocation and multiple access \cite{CM_2021_Cao_AI,Alexandropoulos2022Pervasive,massiveAccessRIS_2022}. The joint optimization of coexisting distributed active and passive nodes in wireless networks also increases the computational burden and control signaling overhead~\cite{TVT_2020_An_Optimal}. Hence, we embark on investigating HMIMO communications by intrinsically integrating programmable metasurfaces with the transceiver.
 \item Over the past decade, we have witnessed the rise of deep learning (DL) techniques. Although DL has been widely utilized for improving the performance of wireless networks, the implementation of DL relies essentially on a central processing unit (CPU) or a graphical processing unit (GPU) \cite{Nature_2015_LeCun_Deep}. Explicitly, DL is merely a computing architecture, whose processing speed is fundamentally constrained by the specific CPU/GPU utilized. To further improve computational efficiency and reduce power consumption, a novel \emph{wave-based computing} paradigm has recently enjoyed much research attention \cite{Science_2018_Lin_All}. Specifically, by constructing a diffractive neural network having a well-designed multilayer structure, the computational tasks can be performed on the profile of the EM wave by leveraging the amplitude/phase information \cite{NCOM_2020_Yao_Protonic}. By fully harnessing the benefits of this wave-based computing paradigm, it becomes possible to perform massively parallel signal processing in the native EM wave regime, where the forward propagation within the diffractive neural network can be realized at the speed of light.
\end{enumerate}
\subsection{Contributions}
Motivated by the aforementioned observations, in this paper, we present an SIM-enabled HMIMO communication system. The main contributions of this paper are summarized as follows:
\begin{enumerate}
 \item We establish a novel HMIMO framework by harnessing an SIM at the transmitter (TX) and another one at the receiver (RX) for achieving substantial spatial gains. The proposed SIM-based transceivers can directly perform precoding/combining in the native EM wave regime with reduced the number of transmit/receive RF chains, which is attributed to the large metasurface aperture.
 \item We formulate a channel fitting problem which focuses on approximating an end-to-end diagonal channel matrix by optimizing the phase shifts associated with the different metasurface layers. This allows each spatial stream to be radiated and recovered independently at the corresponding transmit and receive ports, thus effectively creating a set of interference-free parallel subchannels. By taking into account the constant-modulus constraint and the coupled variables in the objective function, we then propose an efficient gradient descent algorithm for iteratively solving the resultant non-convex problem.
 \item We theoretically analyze the HMIMO channel capacity. Since it is non-trivial to derive the closed-form capacity expression, we provide both an upper and lower bound of the HMIMO capacity by assuming that all the spatial streams experience the best and worst sub-channel quality, respectively. Furthermore, we provide fundamental insights into the scaling law of the HMIMO channel capacity versus the number of data streams and meta-atoms.
 \item Our extensive results demonstrate the benefits of the SIM-aided HMIMO framework conceived as well as the accuracy of our analytical results. We also quantify the channel fitting performance as well as the channel capacity attained under various setups, shedding light on the optimal SIM design. Additionally, we verify the substantial performance improvements attained compared to the conventional MIMO schemes as well as to their RIS-aided counterparts.
\end{enumerate}
\subsection{Organization}
The rest of this paper is organized as follows. In Section \ref{sec2}, we introduce the general SIM-based HMIMO system model, which encompasses the SIM structure together with a spatially correlated HMIMO channel model. Following this, we formulate a channel fitting problem in Section \ref{sec31} and propose an efficient algorithm to address the resulting optimization problem in Section \ref{sec32}. We analyze the HMIMO channel capacity and the computational complexity of the proposed algorithm in Section \ref{sec4}. Finally, our numerical results are provided in Section \ref{sec6} before concluding the paper in Section \ref{sec7}.
\subsection{Notations}
We adopt bold lowercase and uppercase letters to denote vectors and matrices, respectively; ${\left( \cdot \right)^*}$, ${\left( \cdot \right)^T}$, and ${\left( \cdot \right)^H}$ represent the conjugate, transpose, and Hermitian transpose, respectively; $\left| c \right|$, $\Re\left( c \right)$, and $\Im\left( c \right)$ refer to the magnitude, real part, and imaginary part, respectively, of a complex number $c$; $\left\| \cdot \right\|_{\text{F}}$ is the Frobenius norm; $\mathbb{E}\left( \cdot \right)$ stands for the expectation operation; $\log_{a} \left( \cdot \right)$ is the logarithmic function with base $a$, while $\ln \left( \cdot \right)$ is the natural logarithm; $\textrm{diag}\left ( \mathbf{v} \right )$ produces a diagonal matrix with the elements of $\mathbf{v}$ on the main diagonal; $ \mathbf{S}^{1/2}$ denotes the square root of a square matrix $\mathbf{S}$; $\text{vec}\left ( \mathbf{M} \right )$ denotes the vectorization of a matrix $\mathbf{M}$; $\mathbf{M}_{a:b,\, :},\, \mathbf{M}_{:,\, c:d},\, \mathbf{M}_{a:b,\, c:d}$ represent the matrices constructed by extracting $a$-to-$b$-th rows, $c$-to-$d$-th columns, as well as both $a$-to-$b$-th rows and $c$-to-$d$-th columns, respectively, of the matrix $\mathbf{M}$; $\textrm{sinc}\left ( x \right )=\sin\left ( \pi x \right )/\left (\pi x \right )$ is the sinc function; ${{\mathbb{C}}^{x \times y}}$ represents the space of $x \times y$ complex-valued matrices; $\left \lceil x \right \rceil$ refers to the nearest integer greater than or equal to $x$; $\mod \left ( x,n \right )$ returns the remainder after division of $x$ by $n$; $\partial f/\partial x$ means the partial derivative of a function $f$ with respect to (\emph{w.r.t.}) the variable $x$; ${\bf{0}}$ and ${\bf{1}}$ denote all-zero and all-one vectors, respectively, with appropriate dimensions, while ${{\bf{I}}_N} \in \mathbb{C}^{N \times N}$ denotes the identity matrix; the distribution of a circularly symmetric complex Gaussian (CSCG) random vector with mean vector ${\boldsymbol{\mu }}$ and covariance matrix ${\boldsymbol{\Sigma }} \succeq \mathbf{0}$ is denoted by $ \sim {\mathcal{CN}}\left( {{\boldsymbol{\mu }},{\boldsymbol{\Sigma }}} \right)$, where $ \sim$ stands for ``distributed as''.

\begin{figure*}[!t]
\centering
\includegraphics[width=16cm]{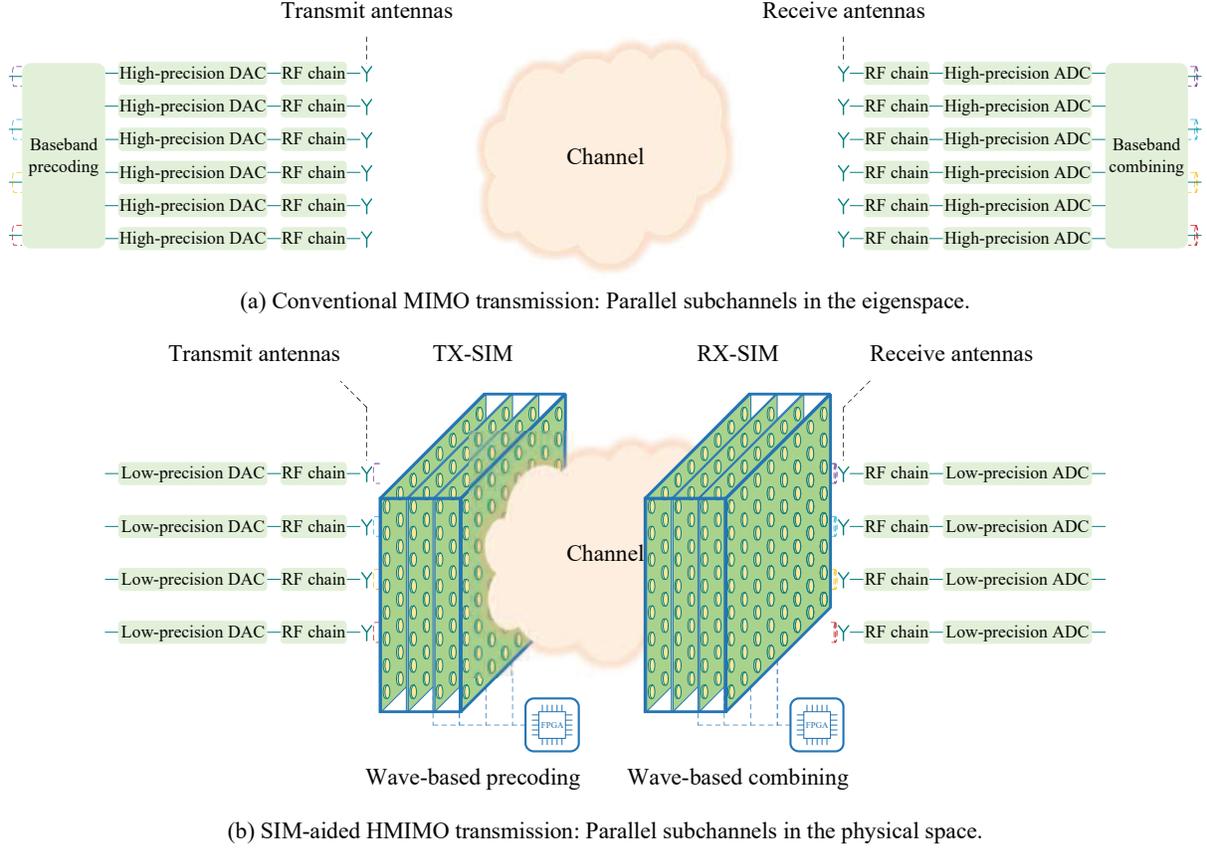}
\caption{Transmission comparison of conventional MIMO and SIM-based HMIMO.}
\label{fig1}
\end{figure*}
\section{The Proposed SIM-Based HMIMO System Model}\label{sec2}
In this section, we present the holistic system model of our SIM-based HMIMO. Specifically, we first introduce the proposed TX/RX SIM-based design and then elaborate on the spatially correlated HMIMO channel model based on the recent consolidated efforts in \cite{WCL_2022_Demir_Channel}. Finally, we discuss the optimal transmission regime of the HMIMO channel, given a limited number of information data streams.
\subsection{Proposed SIM Design}
Before proceeding, let us briefly review the conventional MIMO scheme illustrated in Fig. \ref{fig1}(a), where multiple data streams are first precoded and then fed to the corresponding transmit antennas. At the output of the wireless channel, receiver combining is adopted for recovering the different spatial streams. As a consequence, multiple parallel subchannels are constructed in the eigenspace domain \cite{Book_2005_Tse_Fundamentals}, benefiting from the precoding and combining at the transmitter and receiver, respectively.

The proposed SIM-assisted HMIMO system is illustrated in Fig. \ref{fig1}(b). In sharp contrast to the conventional MIMO systems having only active antennas, an SIM is integrated with both the TX and RX for enhancing the QoS\footnote{Although, in this paper, we only consider the point-to-point HMIMO scenario, our SIM can also be utilized to perform the zero-forcing (ZF) precoding and combining for supporting multiuser HMIMO communications \cite{JSTSP_2022_Wei_Multi, JSAC_2022_Deng_HDMA}. The specific design is beyond the scope of this paper and reserved for our future research.}. Specifically, a TX/RX-SIM is a closed vacuum container having several stacked metasurface layers \cite{NE_2022_Liu_A}. Each metasurface is comprised of a large number of meta-atoms \cite{Light_2014_Cui_Coding}, which are connected to a smart controller, e.g., a field programmable gate array (FPGA) board. By appropriately tuning the drive level of the control circuit associated with each meta-atom, the system becomes capable of manipulating the EM behavior of the penetrating wave, and thus, producing a customized spatial waveform shape at the output of the metasurface layer. Moreover, by compactly arranging large numbers of meta-atoms on the output metasurface of the TX-SIM as well as on the input metasurface of the RX-SIM, the desired information-bearing EM waves can be radiated from almost the entire surface into the ether and then collected in the same way. As such, both the TX-SIM and RX-SIM interact with the wireless channel in an almost continuous manner, thus, being capable to support low-latency \emph{HMIMO communications}. Specifically, the TX-SIM undertakes the precoding task, casting the appropriate information-bearing EM wave into the ether, while the RX-SIM efficiently combines the impinging EM wave on the input surface for signal recovery. As a result, we are able to establish multiple parallel subchannels in the physical space, and the corresponding multiple data streams can be directly radiated and recovered from the associated TX and RX metasurfaces, respectively, without imposing any interference.

\begin{figure*}[!t]
\centering
\includegraphics[width=15cm]{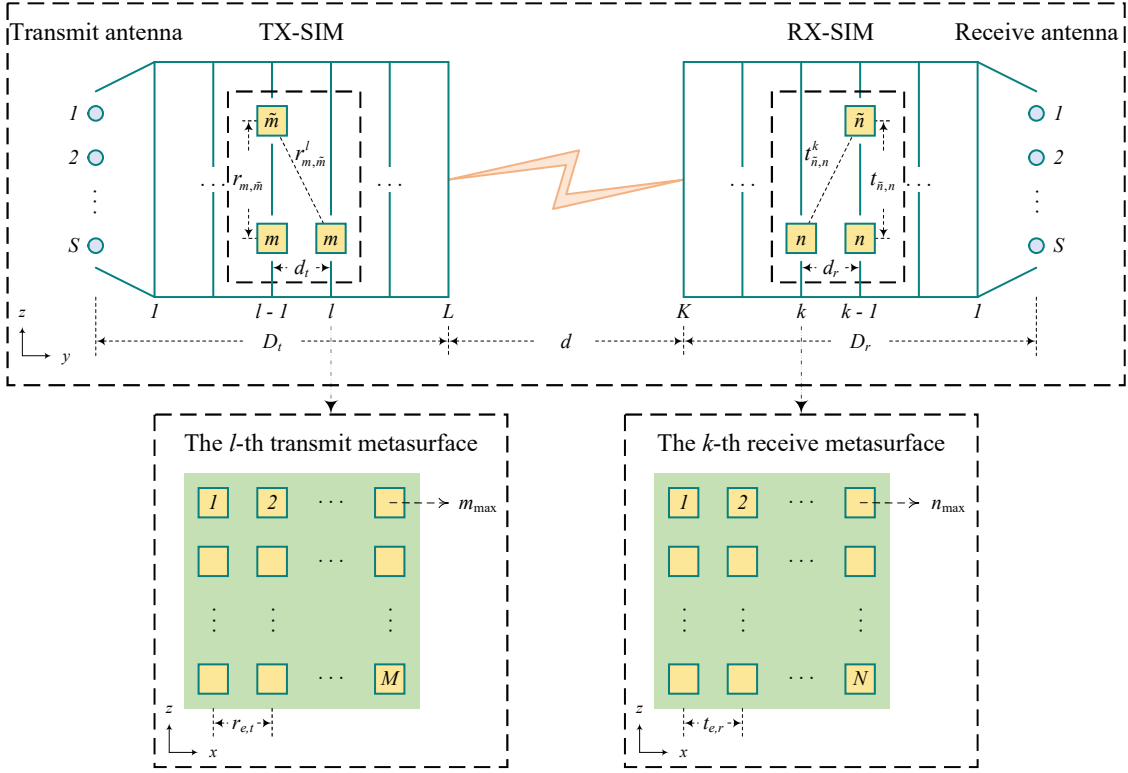}
\caption{Detailed schematic of the SIM-aided HMIMO system of Fig. \ref{fig1}(b).}
\label{fig2}
\end{figure*}

\emph{Remark 1:} Here we elaborate on three core benefits of the proposed SIM-based HMIMO transmission paradigm, as compared to its conventional counterpart \cite{Book_2005_Tse_Fundamentals}. Firstly, the conventional MIMO design requires a large number of active components to achieve spatial gains, thus resulting in high hardware costs and energy consumption. In contrast, the proposed SIM-based HMIMO utilizes low-cost metasurfaces for gleaning spatial gains, which substantially reduces the number of active RF chains required. Secondly, the conventional MIMO transmission solution relies on high-precision power-thirsty digital-to-analog converters (DAC) and analog-to-digital converters (ADC). Instead, the TX/RX-SIM creates multiple parallel subchannels in the physical space, enabling each data stream to be individually processed by low-precision power-efficient DACs/ADCs. For example, $1$-bit resolution may be used for binary phase shift keying (BPSK) without unduly compromising its communication performance. Thirdly, due to using precoding and combining in the wave domain, the power consumption of signal processing is significantly reduced. As such, the SIM substantially reduces the overall energy consumption compared to conventional digital transceiver designs \cite{NE_2022_Liu_A}. Nevertheless, a quantitative evaluation of the energy efficiency of the proposed SIM relying on passive metasurfaces requires an accurate energy consumption model, as well as an accurate transmission model for characterizing the wave propagation between adjacent metasurfaces, both of which require further investigation.

A detailed schematic of the proposed SIM-based HMIMO system, which relies on wave-based precoding and combining, is provided in Fig.~\ref{fig2}. Let $S$ and $\mathcal{S} = \left \{ 1,2,\cdots ,S \right \}$ denote the number of data streams and the corresponding set, respectively. Moreover, $L$ and $K$ denote the number of metasurface layers at the TX and RX, respectively, while their corresponding sets are represented by $\mathcal{L} = \left \{ 1,2,\cdots ,L \right \}$ and $\mathcal{K} = \left \{ 1,2,\cdots ,K \right \}$. For notational convenience, we assume that the number of meta-atoms on each metasurface layer of the TX-SIM is identical and so is for the RX-SIM of Fig. \ref{fig2}. Specifically, let $M$ and $N$ denote the number of meta-atoms on each metasurface layer associated with the TX-SIM and the RX-SIM, respectively, satisfying $M\geq S$ and $N\geq S$, while representing the corresponding set as $\mathcal{M} = \left \{ 1,2,\cdots ,M \right \}$ and $\mathcal{N} = \left \{ 1,2,\cdots ,N \right \}$. Moreover, let $\phi_{m}^{l} =e^{j\theta _{m}^{l}}$ denote the transmission coefficient imposed by the $m$-th meta-atom on the $l$-th TX metasurface layer with $\theta _{m}^{l}$ representing the corresponding phase shift, which we assume that it can be continuously adjusted in the interval between $0$ and $2\pi$, i.e., $\theta _{m}^{l}\in \left [ 0,2\pi \right ),\, m \in \mathcal{M},\, l \in \mathcal{L}$. Thus, the transmission coefficient vector of the $l$-th TX metasurface layer and its corresponding matrix version are denoted by $\boldsymbol{\phi }^{l}=\left [\phi_{1}^{l},\phi_{2}^{l},\cdots ,\phi_{M}^{l} \right ]^{T}\in \mathbb{C}^{M\times 1}$ and $\boldsymbol{\Phi}^{l} = \textrm{diag}\left ( \boldsymbol{\phi}^{l} \right )\in \mathbb{C}^{M\times M}$, respectively. Similarly, let $\psi_{n}^{k} =e^{j\xi _{n}^{k}}$ denote the transmission coefficient imposed by the $n$-th meta-atom on the $k$-th RX metasurface layer, where $\xi _{n}^{k}$ represents the corresponding phase shift satisfying $\xi _{n}^{k}\in \left [ 0,2\pi \right ),\, n \in \mathcal{N},\, k \in \mathcal{K}$. Then, the transmission coefficient vector of the $k$-th RX metasurface layer and its corresponding matrix version are respectively denoted by $\boldsymbol{\psi }^{k}=\left [\psi_{1}^{k},\psi_{2}^{k},\cdots ,\psi_{N}^{k} \right ]^{T}\in \mathbb{C}^{N\times 1}$ and $\boldsymbol{\Psi}^{k} = \textrm{diag}\left ( \boldsymbol{\psi}^{k} \right )\in \mathbb{C}^{N\times N}$.

Furthermore, we assume that all the metasurface layers rely on an isomorphic lattice arrangement \cite{NE_2022_Liu_A}, while each metasurface is modeled as a uniform planar array. Specifically, the element spacing between the $\tilde{m}$-th meta-atom and the $m$-th one on the same TX metasurface and that between the $n$-th meta-atom and the $\tilde{n}$-th one on the same RX metasurface can be expressed as
\begin{align}
 r_{m,\tilde{m}}&=r_{e,t}\sqrt{\left ( m_{z}-\tilde{m}_{z} \right )^{2}+\left ( m_{x}-\tilde{m}_{x} \right )^{2}},\\
 t_{\tilde{n},n}&=t_{e,r}\sqrt{\left ( \tilde{n}_{z}-n_{z} \right )^{2}+\left ( \tilde{n}_{x}-n_{x} \right )^{2}},
\end{align}
respectively, where $r_{e,t}$ and $t_{e,r}$ denote the element spacing between adjacent meta-atoms on the same TX metasurface and that on the same RX metasurface, respectively (see Fig. \ref{fig2}). Additionally, $m_{z}$ and $m_{x}$ denote the indices of the $m$-th meta-atom along the $z$-axis and the $x$-axis, respectively, while $n_{z}$ and $n_{x}$ denote the indices of the $n$-th meta-atom along the $z$-axis and the $x$-axis, respectively, which are defined by
\begin{align}
 m_{z}&=\left \lceil m/m_{\max} \right \rceil,& m_{x}&=\text{mod}\left ( m-1,m_{\max} \right )+1,\\
 n_{z}&=\left \lceil n/n_{\max} \right \rceil,& n_{x}&=\text{mod}\left ( n-1,n_{\max} \right )+1,
\end{align}
respectively, with $m_{\max}$ and $n_{\max}$ denoting the maximum number of meta-atoms on each row of the TX metasurface and that of the RX metasurface, respectively, as shown in Fig. \ref{fig2}. Throughout this paper, we consider square metasurface arrays associated with $M = m_{\max}^2$ and $N = n_{\max}^2$.

Let us now consider the transmission process between the adjacent metasurface layers. For the sake of simplicity, we assume that all the metasurface layers are parallel and have uniform spacing, as shown in Fig. \ref{fig2}. Specifically, let $d_t$ and $d_r$ denote the spacing between any two adjacent metasurface layers in the TX-SIM and that in the RX-SIM, respectively, while $D_t$ and $D_r$ represent the thickness of the TX-SIM and RX-SIM, respectively. Thus we have $d_{t}=D_{t}/L$ and $d_{r}=D_{r}/K$. As a result, the transmission distance from the $\tilde{m}$-th meta-atom on the $\left (l-1 \right )$-st TX metasurface to the $m$-th one on the $l$-th TX metasurface and that from the $n$-th meta-atom of the $k$-th RX metasurface to the $\tilde{n}$-th one on the $\left (k-1 \right )$-st RX metasurface are
\begin{align}
 r_{m,\tilde{m}}^{l}&=\sqrt{r_{m,\tilde{m}}^{2}+d_{t}^{2}},\ l \in \mathcal{L}/\left \{ 1 \right \},\\
 t_{\tilde{n},n}^{k}&=\sqrt{t_{\tilde{n},n}^{2}+d_{r}^{2}},\ k \in \mathcal{K}/\left \{ 1 \right \},
\end{align}
respectively.

Next, we consider the transmission process from the transmit antenna array to the input metasurface of the TX-SIM and that from the output metasurface of the RX-SIM to the receive antenna array. The transmit and receive antennas are both arranged in a uniform linear array, with the element spacing of half-wavelength, i.e., $\lambda /2$, and the array centers aligned with those of all metasurfaces. By doing so, the transmission distance from the $s$-th source to the $m$-th meta-atom on the input metasurface of the TX-SIM and that from the $n$-th meta-atom on the output metasurface of the RX-SIM to the $s$-th destination is given by \eqref{eq7} and \eqref{eq8}, respectively, as shown at the top of this page.

\begin{figure*}
\begin{align}
 r_{m,s}^{1}&=\sqrt{\left [ \left ( m_{z}-\frac{m_{\max}+1}{2} \right )r_{e,t}-\left ( s-\frac{S+1}{2} \right )\frac{\lambda}{2} \right ]^{2}+\left ( m_{x}-\frac{m_{\max}+1}{2} \right )^{2}r_{e,t}^2+d_{t}^{2}}, \label{eq7}\\
 t_{s,n}^{1}&=\sqrt{\left [ \left ( n_{z}-\frac{n_{\max}+1}{2} \right )t_{e,r}-\left ( s-\frac{S+1}{2} \right )\frac{\lambda}{2} \right ]^{2}+\left ( n_{x}-\frac{n_{\max}+1}{2} \right )^{2}t_{e,r}^2+d_{r}^{2}}. \label{eq8}
\end{align}
\hrulefill
\end{figure*}

According to the Rayleigh-Sommerfeld diffraction theory \cite{Science_2018_Lin_All}, the transmission coefficient from the $\tilde{m}$-th meta-atom on the $\left (l-1 \right )$-st TX metasurface layer to the $m$-th meta-atom on the $l$-th TX metasurface layer is expressed by
\begin{align}
 w_{m,\tilde{m}}^{l}=\frac{A_{t}\cos\chi _{m,\tilde{m}}^{l} }{r_{m,\tilde{m}}^{l}}\left ( \frac{1}{2\pi r_{m,\tilde{m}}^{l}}-j\frac{1}{\lambda } \right )e^{j2\pi r_{m,\tilde{m}}^{l}/\lambda },\ l \in \mathcal{L}, \label{eq1}
\end{align}
where $r_{m,\tilde{m}}^{l}$ denotes the corresponding transmission distance, $A_{t}$ is the area of each meta-atom in the TX-SIM, while $\chi _{m,\tilde{m}}^{l}$ represents the angle between the propagation direction and the normal direction of the $\left ( l-1 \right )$-th TX metasurface layer. Let $\mathbf{W}^{l}\in \mathbb{C}^{M\times M},\, l\in \mathcal{L}/\left \{ 1 \right \}$ denote the transmission coefficient matrix between the $\left ( l-1 \right )$-st TX metasurface layer and the $l$-th TX metasurface layer. In particular, the transmission coefficient matrix from the transmit antenna array to the input metasurface layer of the TX-SIM is represented by $\mathbf{W}^{1}\in \mathbb{C}^{M\times S}$. Thus, the effect of the TX-SIM in Fig. \ref{fig2} is formulated by
\begin{align}
 \mathbf{P}=\mathbf{\Phi }^{L}\mathbf{W}^{L}\cdots \mathbf{\Phi }^{2}\mathbf{W}^{2}\mathbf{\Phi }^{1}\mathbf{W}^{1}\in \mathbb{C}^{M\times S}.
\end{align}

Moreover, the transmission coefficient from the $n$-th meta-atom on the $k$-th RX metasurface layer to the $\tilde{n}$-th meta-atom on the $\left (k-1 \right )$-st RX metasurface layer is expressed by \cite{Science_2018_Lin_All}
\begin{align}
 u_{\tilde{n},n}^{k}=\frac{A_{r}\cos\varsigma _{\tilde{n},n}^{k} }{t_{\tilde{n},n}^{k}}\left ( \frac{1}{2\pi t_{\tilde{n},n}^{k}}-j\frac{1}{\lambda } \right )e^{j2\pi t_{\tilde{n},n}^{k}/\lambda },\ k \in \mathcal{K}, \label{eq3}
\end{align}
where $t_{\tilde{n},n}^{k}$ denotes the corresponding transmission distance, $A_{r}$ is the area of each meta-atom in the RX-SIM, while $\varsigma _{\tilde{n},n}^{k}$ represents the angle between the propagation direction and the normal direction of the $\left ( k-1 \right )$-th RX metasurface layer. Let $\mathbf{U}^{k}\in \mathbb{C}^{N\times N},\, k\in \mathcal{K}/\left \{ 1 \right \}$ represent the transmission coefficient matrix between the $k$-th RX metasurface layer to the $\left (k-1 \right )$-st RX metasurface layer, while the transmission coefficient matrix from the output metasurface layer of the RX-SIM to the receive antenna array is denoted by $\mathbf{U}^{1}\in \mathbb{C}^{S\times N}$. Hence, the effect of the RX-SIM in Fig. \ref{fig2} is represented by
\begin{align}
 \mathbf{Q}=\mathbf{U}^{1}\boldsymbol{\Psi }^{1}\mathbf{U}^{2}\boldsymbol{\Psi }^{2}\cdots \mathbf{U}^{K}\boldsymbol{\Psi }^{K}\in \mathbb{C}^{S\times N}.
\end{align}

\emph{Remark 2:} In order to maximize the energy efficiency and characterize the performance of SIM-aided HMIMO communications, we have applied the constant modulus constraint and assumed continuously-adjustable phase shifts for the transmission coefficients associated with each meta-atom \cite{NCOM_2018_Arbabi_MEMS}. While the tuning precision in practice is typically proportional to the hardware costs, the low-resolution meta-atoms may be used in practical SIM design. The specific SIM optimization and the resultant performance analysis taking into account these hardware constraints such as realistic discrete phase shifts \cite{NE_2022_Liu_A}, adjustable magnitudes \cite{CM_2020_Wu_Towards}, as well as coupled phase and magnitude tuning mechanisms \cite{WC_2022_An_Codebook} deserve future exploration.

\emph{Remark 3:} It should be noted that the transmission coefficients between adjacent metasurface layers may deviate from those specified in \eqref{eq1} and \eqref{eq3} due to the existence of practical hardware imperfections, irreversible fabrication shortcomings, as well as innate modeling errors \cite{NE_2022_Liu_A}. Hence, it may be necessary to calibrate these transmission coefficients before the SIM's practical deployment, which can be carried out separately for each individual SIM. One effective method is to transmit a known excitation signal and measure the response at the receive panel, and then update the transmission coefficients by employing the classic error back-propagation algorithm \cite{Nature_2015_LeCun_Deep}. As an initial exploration of the precoding and combining capability of SIM, the specific calibrate process is beyond the scope of this paper and reserved for our future research.

\subsection{Spatially-Correlated HMIMO Channel Model}
Next, let us consider the HMIMO channel model between the TX-SIM and the RX-SIM, where the most prominent property is the spatial correlation among the tightly packed meta-atoms. Specifically, the spatially-correlated HMIMO channel spanning from the output metasurface of the TX-SIM to the input metasurface of the RX-SIM is written by \cite{CL_2022_Hu_Holographic}
\begin{align}
\mathbf{G}=\mathbf{R}_{\textrm{Rx}}^{1/2}\tilde{\mathbf{G}}\mathbf{R}_{\textrm{Tx}}^{1/2}\in \mathbb{C}^{N\times M}, \label{eq5}
\end{align}
where $\tilde{\mathbf{G}} \in \mathbb{C}^{N\times M}$ is the independent and identically distributed (i.i.d.) Rayleigh fading channel, i.e., $\tilde{\mathbf{G}} \sim \mathcal{CN}\left ( \mathbf{0},\rho ^{2}\mathbf{I}_{N}\otimes \mathbf{I}_{M} \right )$ with $\rho ^{2}$ denoting the average path loss between the pair of wireless transceivers, while $\mathbf{R}_{\textrm{Tx}}\in \mathbb{C}^{M\times M}$ and $\mathbf{R}_{\textrm{Rx}}\in \mathbb{C}^{N\times N}$ represent the spatial correlation matrix at the TX-SIM and that at the RX-SIM, respectively. By considering far-field propagation in an isotropic scattering environment \cite{JSAC_2020_Pizzo_Spatially, Access_2020_Dai_Reconfigurable}, the spatial correlation matrix at the TX-SIM and that at the RX-SIM can be expressed by \cite{WCL_2022_Demir_Channel}
\begin{align}
 \left [ \mathbf{R}_{\text{Tx}} \right ]_{m,\tilde{m}}&=\text{sinc}\left ( 2r_{m,\tilde{m}}/\lambda \right ),\ \tilde{m} \in \mathcal{M},\ m \in \mathcal{M},\\
 \left [ \mathbf{R}_{\text{Rx}} \right ]_{\tilde{n},n}&=\text{sinc}\left ( 2t_{\tilde{n},n}/\lambda \right ),\ n \in \mathcal{N},\ \tilde{n} \in \mathcal{N},
\end{align}
respectively.

Moreover, the path loss between the transmitter and the receiver is modeled by \cite{TCOM_2015_Rappaport_Wideband}
\begin{align}
 \text{PL}\left ( d \right )=\text{PL}\left ( d_{0} \right )+10b\log_{10}\left ( \frac{d}{d_{0}} \right )+X_{\delta },\ d\geq d_{0}, \label{eq16}
\end{align}
where $\text{PL}\left ( d_{0} \right ) = 20\log_{10}\left ( 4\pi d_{0}/\lambda \right )$ dB is the free space path loss at the reference distance $d_0$, $b$ represents the path loss exponent, $X_{\delta }$ is a zero mean Gaussian random variable with a standard deviation $\delta$, characterizing the large-scale signal fluctuations of shadow fading.

\emph{Remark 4:} Note that the spatial correlation matrix highly depends on the scattering environments surrounding both the transmitter and the receiver. Therefore, it is generally unrealistic to derive a universal spatially-correlated fading model that can be applied to all practical communication scenarios. Motivated readers are referred to \cite{JSAC_2020_Pizzo_Spatially, TWC_2022_Pizzo_Fourier, WCL_2022_Demir_Channel} and references therein for gaining further insights concerning the channel models for HMIMO systems that are derived from the intrinsic EM propagation properties.

\subsection{SIM-Aided HMIMO Channel Capacity with Limited Number of Streams}
In this subsection, we will consider both the HMIMO channel capacity as well as the optimal transmission given a limited number of data streams. Specifically, given the wireless channel $\mathbf{G}$ and the fixed number of data streams $S$, the optimal HMIMO transmission is achieved by applying the truncated singular value decomposition (SVD) policy \cite{TCOM_2022_An_Low}. The detailed procedures are outlined as follows:
\begin{itemize}
 \item[\emph{1:}] First of all, we perform the SVD of $\mathbf{G}$ so that $\mathbf{G}=\mathbf{E}\mathbf{\Lambda }\mathbf{F}^{H}$, where we have $\mathbf{\Lambda }=\textrm{diag}\left ( \lambda _{1},\lambda _{2},\cdots ,\lambda _{O} \right )$ and $O=\min\left ( M,N \right )$, while $\lambda _{1}\geq \lambda _{2}\geq \cdots \geq \lambda _{O}$ denoting the singular values in non-increasing order.
 \item[\emph{2:}] Next, by spreading the data streams using a transmit precoder $\mathbf{P}=\mathbf{F}_{:,1:S}\in \mathbb{C}^{M\times S}$ and collecting the spatial signals using a receive combiner $\mathbf{Q}=\mathbf{E}_{:,1:S}^{H}\in \mathbb{C}^{S\times N}$, the resultant end-to-end channel becomes the following diagonal matrix
 \begin{align}
 \mathbf{H}=\mathbf{Q}\mathbf{G}\mathbf{P} = \mathbf{\Lambda} _{1:S,1:S}\in \mathbb{C}^{S\times S},
 \end{align}
with $\mathbf{\Lambda} _{1:S,1:S}$ being the $S$-th order leading principal minor of $\mathbf{\Lambda}$.
 \item[\emph{3:}] Furthermore, in order to maximize the channel capacity, the optimal power allocation coefficients can be obtained by applying the well-known water-filling algorithm \cite{WCL_2022_An_Scalable}. Specifically, the amount of power allocated to the $s$-th data stream is determined as
\begin{align}
 p_{s}=\max\left ( 0, \tau -\frac{\sigma ^{2}}{\lambda _{s}^{2}}\right ),\ s\in \mathcal{S}, \label{eq9}
\end{align}
where $\tau$ is a threshold value satisfying the total transmit power constraint, i.e., $\sum\nolimits_{s=1}^{S}p_{s}=P_{t}$ with $P_{t}$ denoting the total available power at the transmitter, which can be obtained by utilizing the bisection method, while $\sigma ^{2}$ represents the average noise power at the receiver.
\item[\emph{4:}] Therefore, the HMIMO channel capacity for a finite number of data streams is given by
\begin{align}
 C=\sum_{s=1}^{S}\log_{2}\left ( 1+\frac{p_{s}\lambda _{s}^{2}}{\sigma ^{2}} \right ). \label{eq10}
\end{align}
\end{itemize}

Next, let us get back to our SIM-aided HMIMO communication system shown in Fig. \ref{fig2}. In sharp contrast to conventional MIMO designs constructing multiple virtual streams in the eigenspace by employing the appropriate digital precoding and combining, we endeavor to naturally form multiple parallel physical subchannels between the transmit antennas and their corresponding receive antennas. Explicitly, the precoding and combining are implemented in the EM wave regime by optimizing the TX-SIM and RX-SIM as follows
\begin{align}
 \mathbf{F}_{:,1:S}&\simeq \mathbf{\Phi }^{L}\mathbf{W}^{L}\cdots \mathbf{\Phi }^{2}\mathbf{W}^{2}\mathbf{\Phi }^{1}\mathbf{W}^{1},\\
 \mathbf{E}_{:,1:S}^{H}&\simeq \mathbf{U}^{1}\boldsymbol{\Psi }^{1}\mathbf{U}^{2}\boldsymbol{\Psi }^{2}\cdots \mathbf{U}^{K}\boldsymbol{\Psi }^{K}.
\end{align}
Thus, one might be able to construct an end-to-end diagonal channel $\mathbf{H}$, such that multiple data streams can be directly radiated and recovered, respectively, from the corresponding transmit and receive antennas. In a nutshell, SIM not only reaps spatial gains benefiting from the massive number of meta-atoms on the metasurface layer but performs the precoding and combining at the speed of light, thanks to its direct wave-based computing capability attained by the multilayer structure.
\section{Problem Formulation and Solution of Joint Optimizing TX-SIM and RX-SIM}\label{sec3}
\subsection{Problem Formulation}\label{sec31}
In this subsection, we formulate the problem of minimizing the error between the end-to-end channel of $\mathbf{H}=\mathbf{Q}\mathbf{G}\mathbf{P}$ and the expected diagonal matrix $\boldsymbol{\Lambda}_{1:S,1:S}$ by optimizing the phase shifts of the TX-SIM and RX-SIM in Fig. \ref{fig2}. We adopt the Frobenius norm to characterize the fitting error of the desired channel fitting problem \cite{JSTSP_2016_Heath_An}. Specifically, the optimization problem is formulated as\footnote{Note that although the formulated problem seems to have a similar form to that in hybrid beamforming, e.g., \cite{JSTSP_2016_Heath_An}, they are fundamentally different because the proposed SIM-assisted HMIMO fully removes the digital precoding and combining by utilizing the multilayer metasurface structure, while imposing the phase shifts in the wave regime.}
\begin{subequations}\label{eq13}
\begin{alignat}{2}
&{\underset{\phi _{m}^{l},\, \psi _{n}^{k},\, \alpha}{\text{minimize} }} \ &&{\Gamma = \left \| \alpha \mathbf{Q}\mathbf{G}\mathbf{P}-\boldsymbol{\Lambda}_{1:S,1:S} \right \|_{\text{F}}^{2}} \label{eq13a}\\
&{\text{subject\ to}} \ && \mathbf{P}=\mathbf{\Phi }^{L}\mathbf{W}^{L}\cdots \mathbf{\Phi }^{2}\mathbf{W}^{2}\mathbf{\Phi }^{1}\mathbf{W}^{1},\\
&{} \ &&\mathbf{Q}=\mathbf{U}^{1}\boldsymbol{\Psi }^{1}\mathbf{U}^{2}\boldsymbol{\Psi }^{2}\cdots \mathbf{U}^{K}\boldsymbol{\Psi }^{K},\\
&{} \ &&\boldsymbol{\Phi}^{l} = \textrm{diag}\left ( \left [\phi_{1}^{l},\phi_{2}^{l},\cdots ,\phi_{M}^{l} \right ]^{T} \right ),\ l\in \mathcal{L},\\
&{} \ &&\boldsymbol{\Psi}^{k} = \textrm{diag}\left ( \left [\psi_{1}^{k},\psi_{2}^{k},\cdots ,\psi_{N}^{k} \right ]^{T} \right ),\ k\in \mathcal{K},\\
&{} \ &&{\left | \phi _{m}^{l} \right | = 1},\ m \in \mathcal{M},\ l \in \mathcal{L}, \label{eq13d}\\
&{} \ &&{\left | \psi _{n}^{k} \right | = 1},\ n \in \mathcal{N},\ k \in \mathcal{K}, \label{eq13e}\\
&{} \ &&{\alpha \in \mathbb{C}}, \label{eq13f}
\end{alignat}
\end{subequations}
where $\alpha$ is a scaling factor compensated by SIM. 

\emph{Remark 5:} Our objective is to evaluate the signal processing capability of SIM having multiple stacked metasurface layers. As such, we have assumed that the TX-SIM and RX-SIM of Fig. \ref{fig2} compensate for an adaptive gain $\alpha$, as seen in \eqref{eq13a}. Although this assumption might seem somewhat simplified, it is reasonable due to the fact that conventional precoding and combining architectures relying on digital signal processing result in additional hardware costs and energy consumption. Thus, a fair performance comparison between these two paradigms requires special attention under the same resource consumption. Since the energy consumption of our proposed SIM is unexplored, it poses an open challenge in conducting a fair performance comparison.

Note that due to the non-convex constant modulus constraint on each transmission coefficient, i.e., \eqref{eq13d} and \eqref{eq13e}, as well as the highly coupled variables in the objective function, i.e., \eqref{eq13a}, it is non-trivial to obtain the optimal solution of Problem \eqref{eq13}. As such, in Section \ref{sec32}, we will provide an efficient algorithm for solving the channel fitting problem iteratively.
\subsection{The Proposed Gradient Descent Algorithm}\label{sec32}
In this section, an efficient gradient descent algorithm is proposed for solving the challenging Problem \eqref{eq13}. To ensure compliance with the constant modulus constraints, i.e., \eqref{eq13d} and \eqref{eq13e}, our gradient descent algorithm is implemented by deriving the partial derivative \emph{w.r.t.} the phase shifts. As such, the constant modulus constraints can be guaranteed throughout the iteration process. The detailed steps of the iteration core of the proposed gradient descent algorithm are summarized as follows.

\emph{\textbf{Step 1: Calculate the partial derivatives}}

First, the partial derivatives of the loss function $\Gamma$ \emph{w.r.t.} the $m$-th phase shift $\theta _{m}^{l}$ of the $l$-th TX metasurface layer and that \emph{w.r.t.} the $n$-th phase shift $\xi _{n}^{k}$ of the $k$-th RX metasurface layer are respectively given by
\begin{align}
 \frac{\partial \Gamma}{\partial \theta _{m}^{l}}&=2\sum_{s=1}^{S}\sum _{\tilde{s}=1}^{S}\Im\left [ \left (\alpha\phi _{m}^{l} x_{m,s,\tilde{s}}^{l} \right )^{\ast}\left ( \alpha h_{s,\tilde{s}} - \lambda _{s,\tilde{s}}\right ) \right ], \label{eq14} \\
 \frac{\partial \Gamma}{\partial \xi _{n}^{k}}&=2\sum_{s=1}^{S}\sum _{\tilde{s}=1}^{S}\Im\left [ \left (\alpha \psi _{n}^{k}y_{n,s,\tilde{s}}^{k} \right )^{\ast}\left ( \alpha h_{s,\tilde{s}} - \lambda _{s,\tilde{s}}\right ) \right ], \label{eq15}
\end{align}
where $h_{s,\tilde{s}}$ and $\lambda _{s,\tilde{s}}$ denote the entries located at the $s$-th row and the $\tilde{s}$-th column of the matrix $\mathbf{H} = \mathbf{Q}\mathbf{G}\mathbf{P}$ and that of the matrix $\boldsymbol{\Lambda}$, respectively; while $x_{m,s,\tilde{s}}^{l}$ and $y_{n,s,\tilde{s}}^{k}$ denote the cascaded channel spanning from the $\tilde{s}$-th transmit antenna to the $s$-th receive antenna via the $m$-th meta-atom of the $l$-th TX metasurface layer and that via the $n$-th meta-atom of the $k$-th RX metasurface layer, which are defined by
\begin{align}
 x_{m,s,\tilde{s}}^{l}&=\mathbf{Q}_{s,:}\mathbf{G}\mathbf{\Phi }^{L}\mathbf{W}^{L}\cdots \mathbf{W}_{:,m}^{l+1}\mathbf{W}_{m,:}^{l}\cdots\mathbf{\Phi }^{1}\mathbf{W}_{:,\tilde{s}}^{1},\\
 y_{n,s,\tilde{s}}^{k}&=\mathbf{U}_{s,:}^{1}\boldsymbol{\Psi }^{1}\cdots\mathbf{U}_{:,n}^{k}\mathbf{U}_{n,:}^{k+1}\cdots \mathbf{U}^{K}\boldsymbol{\Psi }^{K}\mathbf{G}\mathbf{P}_{:,\tilde{s}},
\end{align}
respectively.

\emph{\textbf{Step 2: Normalize the partial derivatives}}

In order to mitigate the potential gradient explosion and vanishing problems \cite{BD_2020_Basodi_Gradient}, we normalize the partial derivatives at each iteration as follows
\begin{align}
 \frac{\partial \Gamma}{\partial {\theta _{m}^{l}}}&\leftarrow\frac{\pi }{\varrho_{l}}\cdot \frac{\partial \Gamma}{\partial {\theta _{m}^{l}}},\ m\in \mathcal{M},\ l\in \mathcal{L}, \label{eq18}\\
 \frac{\partial \Gamma}{\partial {\xi _{n}^{k}}}&\leftarrow \frac{\pi }{\varepsilon_{k} } \cdot\frac{\partial \Gamma}{\partial {\xi _{n}^{k}}},\ n\in \mathcal{N},\ k\in \mathcal{K}, \label{eq19}
\end{align}
where we have $\varrho _{l}=\underset{m\in \mathcal{M}}{\max}\left ( \frac{\partial \Gamma}{\partial {\theta _{m}^{l}}} \right ),\, l \in \mathcal{L}$ and $\varepsilon_{k}=\underset{n\in \mathcal{N}}{\max}\left ( \frac{\partial \Gamma}{\partial {\xi _{n}^{k}}} \right ),\, k \in \mathcal{K}$ denoting the maximum value of the partial derivative associated with the $l$-th TX metasurface layer and that with the $k$-th RX metasurface layer, respectively. Note that the normalization process also has the benefit of allowing us to readily pick an initial learning rate independent of the data value \cite{BD_2020_Basodi_Gradient}.

\emph{\textbf{Step 3: Update the phase shifts}}

Then the phase shifts associated with the TX-SIM and RX-SIM in Fig. \ref{fig2} can be updated by
\begin{align}
 \theta _{m}^{l}&\leftarrow \theta _{m}^{l}-\eta \frac{\partial \Gamma}{\partial {\theta _{m}^{l}}},\ m\in \mathcal{M},\ l\in \mathcal{L}, \label{eq20}\\
 \xi _{n}^{k}&\leftarrow \xi _{n}^{k}-\eta \frac{\partial \Gamma}{\partial {\xi _{n}^{k}}},\ n\in \mathcal{N},\ k\in \mathcal{K}, \label{eq21}
\end{align}
respectively, where $ \eta > 0 $ denotes the learning rate that determines the step size at each iteration.

\emph{\textbf{Step 4: Update the scaling factor}}

Given a set of phase shifts associated with the TX-SIM and RX-SIM, the resultant end-to-end channel becomes $\mathbf{H}=\mathbf{Q}\mathbf{G}\mathbf{P}$. Consequently, the optimal scaling factor can be readily obtained by applying the least-square technique as follows
\begin{align}
\alpha =\left (\mathbf{h }^{H}\mathbf{h } \right )^{-1}\mathbf{h }^{H}\boldsymbol{\lambda }, \label{eq22}
\end{align}
where we have $\boldsymbol{\lambda } = \textrm{vec}\left ( \boldsymbol{\Lambda }_{1:S,1:S} \right ) \in \mathbb{C}^{S^2 \times 1}$ and $\mathbf{h} = \textrm{vec}\left ( \mathbf{H} \right ) \in \mathbb{C}^{S^2 \times 1}$ denoting the vectorization of $\boldsymbol{\Lambda }_{1:S,1:S}$ and that of $\mathbf{H}$, respectively.

\emph{\textbf{Step 5: Update the learning rate}}

For the sake of avoiding any overshooting effects \cite{Nature_2015_LeCun_Deep}, we adopt a negative exponentially decaying learning schedule for decreasing the learning rate, as the iteration proceeds. More specifically, the learning rate is updated by
\begin{align}
 \eta \leftarrow \eta \beta , \label{eq23}
\end{align}
at each iteration, where $0 < \beta < 1$ is a hyperparameter controlling the decay rate.

After repeating \eqref{eq14} $\sim$ \eqref{eq23} several times, the loss function $\Gamma$ gradually approaches convergence. In order to prevent the gradient descent algorithm from getting trapped in a local optimum, we first generate multiple sets of phase shifts and then select the one that minimizes $\Gamma$ for initialization. For clarity, we summarize the detailed procedures of the proposed gradient descent in Table \ref{tab1}.

\begin{table}[!t]
\caption{The proposed gradient descent algorithm for solving \eqref{eq13}.}\label{tab1}
\begin{tabular}{l}
\toprule
\hspace{0.15cm}1: {\textbf{INPUT:}} $\mathbf{W}^{l},\, l\in \mathcal{L}$, $\mathbf{G}$, $\mathbf{U}^{k},\, k\in \mathcal{K}$, $\boldsymbol{\Lambda }_{1:S,1:S}$.\\
\hspace{0.15cm}2: Randomly initializing the phase shifts, i.e., $\boldsymbol{\theta} ^{l},\, l\in \mathcal{L}$, $\boldsymbol{\xi} ^{k},\, k\in \mathcal{K}$;\\
\hspace{0.15cm}3: Calculating the scaling factor $\alpha $ by applying (\ref{eq22});\\
\hspace{0.15cm}4: {\textbf{REPEAT}}\\
\hspace{0.15cm}5: \quad Calculating the partial derivatives of $\Gamma $ \emph{w.r.t.} $\theta _{m}^{l}$ and that \emph{w.r.t.}\\
\hspace{0.75cm}$\xi _{n}^{k}$ by applying (\ref{eq14}) and \eqref{eq15}, respectively;\\
\hspace{0.15cm}6: \quad Normalizing the partial derivatives of $\Gamma $ \emph{w.r.t.} $\theta _{m}^{l}$ and that \emph{w.r.t.}\\
\hspace{0.75cm}$\xi _{n}^{k}$ by applying (\ref{eq18}) and \eqref{eq19}, respectively;\\
\hspace{0.15cm}7: \quad Updating the phase shifts, i.e., $\theta _{m}^{l}$ and $\xi _{n}^{k}$, by applying \eqref{eq20} and\\
\hspace{0.75cm}\eqref{eq21}, respectively;\\
\hspace{0.15cm}8: \quad Updating the scaling factor $\alpha $ by applying (\ref{eq22});\\
\hspace{0.15cm}9: \quad Diminishing the learning rate $\eta$ by applying (\ref{eq23});\\
10: \quad Calculating the objective function value $\Gamma$ by applying (\ref{eq13a});\\
11: {\textbf{UNTIL}} The decrement of $\Gamma $ is less than a preset threshold value or\\
\hspace{1.45cm}the number of iterations reaches the maximum;\\
12: {\textbf{OUTPUT:}} $\boldsymbol{\theta} ^{l},\, l\in \mathcal{L}$, $\boldsymbol{\xi} ^{k},\, k\in \mathcal{K}$.\\
\bottomrule
\end{tabular}
\end{table}
\section{Performance Analysis}\label{sec4}
\subsection{HMIMO Channel Capacity Analysis}\label{sec41}
In this subsection, we evaluate the channel capacity of our HMIMO system. We assume that the TX-SIM and RX-SIM shown in Fig. \ref{fig2} have carried out perfect precoding and combining in the EM regime. However, due to the fact that \eqref{eq10} cannot be readily expressed in closed form, here we provide an upper and a lower bound for the channel capacity of our HMIMO system. Specifically, by assuming that all the data streams experience either the best or the worst subchannel, respectively, the ergodic channel capacity is upper and lower bounded by
\begin{align}
 S\log_{2}\left ( 1+\frac{P_{t}\mathbb{E}\left ( \lambda _{S}^{2} \right )}{S\sigma ^{2}} \right )\leq \mathbb{E}\left ( C \right ) \leq S\log_{2}\left ( 1+\frac{P_{t}\mathbb{E}\left ( \lambda _{1}^{2} \right )}{S\sigma ^{2}} \right ), \label{eq33}
\end{align}
where $\mathbb{E}\left ( \lambda _{1}^{2} \right )$ and $\mathbb{E}\left ( \lambda _{S}^{2} \right )$ denote the statistical average of the $1$-st and the $S$-th eigenvalues, respectively, which are obtained through numerical approximations.

In order to gain some fundamental insights into the HMIMO channel capacity, we next analyze its scaling law by considering some special cases. Specifically, the HMIMO channel capacity evaluated by considering a large number of data streams is characterized by \emph{Proposition 1}.

\emph{Proposition 1:} As $S\rightarrow \infty $, we have $P_{t}\log_{2}e\mathbb{E}\left ( \lambda _{S}^{2} \right )/\sigma ^{2}\leq \mathbb{E}\left ( C \right ) \leq P_{t}\log_{2}e\mathbb{E}\left ( \lambda _{1}^{2} \right )/\sigma ^{2}$.

\emph{Proof:} By taking the limit of lower bound and upper bound in \eqref{eq33} as $S\rightarrow \infty $ and applying the formula $\underset{x\rightarrow \infty }{\lim}\left ( 1+1/x \right )^{x}=e$, the proof is completed. $\hfill\blacksquare$

\emph{Proposition 1} demonstrates that blindly increasing the number of active components may not lead to substantial improvements in channel capacity. Specifically, the HMIMO channel capacity gradually saturates as the number of data streams increases due to the intrinsic limitations of \emph{spatial multiplexing} in the HMIMO channel. Further increasing the number of active components may result in severe energy efficiency degradation. Therefore, the resultant HMIMO channel capacity critically depends on the statistical expectation of the eigenvalues, which results in a \emph{selection gain} associated with the increased number of meta-atoms.

In order to characterize the fundamental scaling law of the HMIMO channel capacity versus the number of meta-atoms, we next consider the particular case of $S = 1$, $L = K = 1$ and assume i.i.d. Rayleigh fading for the sake of brevity. Specifically, the theoretical ergodic channel capacity is summarized in \emph{Proposition 2}.

\emph{Proposition 2:} As $M,\, N \rightarrow \infty $, we have $\mathbb{E}\left ( C \right )\simeq\log_{2}\left ( 1+\frac{\pi^{2} P_{t}\rho ^{2}}{4\sigma ^{2}}M^{2}N^{2} \right )$.

\emph{Proof:} As $M,\, N \rightarrow \infty $, the HMIMO channel capacity can be approximated by
\begin{align}\label{eq30}
 \mathbb{E}\left ( C \right ) \simeq \log_{2}\left [ 1+\frac{P_{t}}{\sigma ^{2}} \mathbb{E}\left ( \left | h \right |^{2} \right ) \right ], \end{align}
with
 \begin{align}
 \mathbb{E}\left ( \left | h \right |^{2} \right )=\rho ^{2}\mathbb{E}\left ( \left | \sum_{m=1}^{M}\phi _{m}h_{1,m} \right |^{2} \left | \sum_{n=1}^{N}\psi _{n}h_{2,n} \right |^{2} \right ),
\end{align}
where $h_{1,m} \sim \mathcal{CN}\left ( 0,1\right ),\, m \in \mathcal{M}$ and $h_{2,n} \sim \mathcal{CN}\left ( 0,1\right ),\, n \in \mathcal{N}$ denote the normalized channel coefficient of the link spanning from the source to the optimal scatterer via the $m$-th transmit meta-atom and that from the optimal scatterer to the destination via the $n$-th receive meta-atom, respectively. Note that the unnecessary superscripts and subscripts have been removed for the sake of brevity.

Furthermore, by applying the optimal phase shift configuration, i.e., $\phi _{m}=h_{1,m}^{\ast }/\left |h_{1,m} \right |$ and $\psi _{n}=h_{2,n}^{\ast }/\left |h_{2,n} \right |$, we have \cite{TCOM_2021_Wu_Intelligent}
\begin{align}
 \mathbb{E}\left ( \left | \sum_{m=1}^{M}\phi _{m}h_{1,m} \right |^{2} \right ) &= \mathbb{E}\left (\left | \sum_{m=1}^{M}\left | h_{1,m} \right | \right |^{2} \right ) = \frac{\pi }{2}M^2, \label{eq31}\\
\mathbb{E}\left ( \left | \sum_{n=1}^{N}\psi _{n}h_{2,n} \right |^{2} \right ) &= \mathbb{E}\left (\left | \sum_{n=1}^{N}\left | h_{2,n} \right | \right |^{2} \right ) = \frac{\pi }{2}N^2. \label{eq32}
\end{align}

By substituting \eqref{eq31} and \eqref{eq32} into \eqref{eq30}, the proof is completed. $\hfill\blacksquare$

\emph{Proposition 2} testifies to the quadratic scaling law of the channel gain versus the number of meta-atoms \cite{TCOM_2021_Wu_Intelligent}. Note that in contrast to the RIS-aided system \cite{TCOM_2021_Wu_Intelligent}, both the TX-SIM and RX-SIM could attain spatial gains. In an ideal setup, one could obtain about 4 bps/Hz HMIMO channel capacity improvement upon every doubling of the number of meta-atoms at both the TX-SIM and RX-SIM. Again, we note that the rigorous capacity analysis of SIM-aided HMIMO systems having an arbitrary number of metasurface layers is a complex task due to the fact that a large number of matrix multiplications are involved during the forward propagation \cite{NE_2022_Liu_A, Science_2018_Lin_All}. To address this issue, effective matrix analysis tools might be employed for evaluating the fitting performance of the proposed SIM as well as the resultant channel capacity, which requires further investigation. Nevertheless, our numerical results of Section \ref{sec6} demonstrate that harnessing a pair of SIMs having a moderate number of metasurface layers at both ends can fit the end-to-end channel with high accuracy. As such, motivated readers may refer to \cite{TWC_2022_Pizzo_Fourier} for gaining deeper insights concerning the HMIMO channel capacity relying on conventional digital precoding and combining, which serves as an upper bound for our SIM-aided HMIMO system.

\subsection{Computational Complexity Analysis}\label{sec42}
Next, we analyze the computational complexity of the proposed gradient descent algorithm in terms of the number of real-valued multiplications. Specifically, the computational complexity of performing Step 1 includes that of performing the forward propagation, i.e., $\mathcal{O}_{1-1} = 4SM\left ( ML-M+L \right )+4SN\left ( NK-N+K \right )+4MSN+2\left ( M+N \right )S^{2}$, and that of calculating all partial derivatives, i.e., $\mathcal{O}_{1-2} = 2\left ( ML+NK \right )S^{2}$. Additionally, the computational complexities of performing Steps 2 $\sim$ 5 are $\mathcal{O}_2 = 2\left ( ML+NK \right )$, $\mathcal{O}_3 = \left ( ML+NK \right )$, $\mathcal{O}_4 = 6S^2+2$, and $\mathcal{O}_5 = 1$, respectively. As a result, the total computational complexity of the proposed gradient descent algorithm is given by
\begin{align}
 \mathcal{O} &=I\left [ \mathcal{O}_{1-1}+\mathcal{O}_{1-2}+\mathcal{O}_{2}+\mathcal{O}_{3}+\mathcal{O}_{4}+\mathcal{O}_{5}\right ] \notag\\
 &=I\left [ 4SM\left ( ML-M+L \right )+4SN\left ( NK-N+K \right ) \right ]\notag\\
 &+I\left [4MSN+2\left ( M+N \right )S^{2} \right ]+2I\left ( ML+NK \right )S^{2} \notag\\
&+3I\left ( ML+NK \right )+ I\left ( 6S^2+3\right ) \notag\\
&\simeq 4IS\left ( M^{2}L+N^{2}K \right ),\quad \text{for}\ M,\, N\gg S, 
\end{align}
where $I$ denotes the number of iterations, which will be shown in Section \ref{sec6} to be less than $20$ under an empirical setup. Thus, the proposed gradient descent algorithm is of polynomial-time complexity, when solving Problem \eqref{eq13}.

\section{Simulation Results}\label{sec6}
In this section, we provide numerical results for characterizing the performance of the proposed SIM-aided HMIMO system.
\subsection{Simulation Setups}
As illustrated in Fig. \ref{fig2}, we consider an SIM-aided HMIMO system. In our simulations, the thicknesses of both the TX-SIM and RX-SIM are set to $D_{t}=D_{r}=0.05$ m. Accordingly, the transmission coefficients between the adjacent metasurface layers in the TX-SIM and that in the RX-SIM are given by \eqref{eq1} and \eqref{eq3}, respectively, while the HMIMO channel is generated by \eqref{eq5}. Our SIM-aided HMIMO system operates at the frequency of $f_0 = 28$ GHz, which corresponds to the wavelength of $\lambda = 10.7$ mm. To account for the large-scale fading, we consider the reference distance of $d_0=1$ m and set $b = 3.5$ and $\delta =9$ dB in our simulations \cite{TCOM_2015_Rappaport_Wideband}. Moreover, the distance between the transmitter and the receiver is set to $d = 250$ m.

Additionally, the total power available at the transmitter is set to $P_t = 20$ dBm, while the receiver sensitivity is set to $\sigma ^{2} = -110$ dBm. For the proposed gradient descent algorithm, the number of randomizations for initialization is set to $10$. The maximum affordable number of iterations is set to $100$, while the initial learning rate and decay parameter are set to $\eta_{0} = 0.1$ and $\beta=0.5$, respectively, unless otherwise specified. All the simulation results are obtained by averaging over $100$ independent experiments. Moreover, we adopt a couple of different performance metrics. Specifically, for the sake of a fair comparison, we first quantify the normalized mean square error (NMSE) between the actual channel matrix and the target diagonal one defined as follows
\begin{align}
 \Delta =\mathbb{E}\left ( \frac{\left \| \alpha \mathbf{Q}\mathbf{G}\mathbf{P}-\boldsymbol{\Lambda}_{1:S,1:S} \right \|_{\text{F}}^{2}}{\left \| \boldsymbol{\Lambda}_{1:S,1:S} \right \|_{\text{F}}^{2}} \right ),
\end{align}
while the other is the channel capacity of our SIM-assisted HMIMO system, which is defined by
\begin{align}
 C=\sum_{s=1}^{S}\log_{2}\left ( 1+\frac{p_{s}\left | \alpha h_{s,s} \right |^{2}}{\sum_{\tilde{s}\neq s}^{S}p_{\tilde{s}}\left | \alpha h_{s,\tilde{s}} \right |^{2}+\sigma ^{2}} \right ), \label{eq39}
\end{align}
where $p_s$ denotes the amount of power allocated to the $s$-th stream obtained by \eqref{eq9}. Note that in sharp contrast to conventional MIMO designs with extra digital combining, our SIM-aided HMIMO treats the residual signals from other data streams as interference, as seen in the denominator of \eqref{eq39}.

\subsection{Performance versus System Parameters}
\begin{figure}[!t]
\centering
\includegraphics[width=7.5cm]{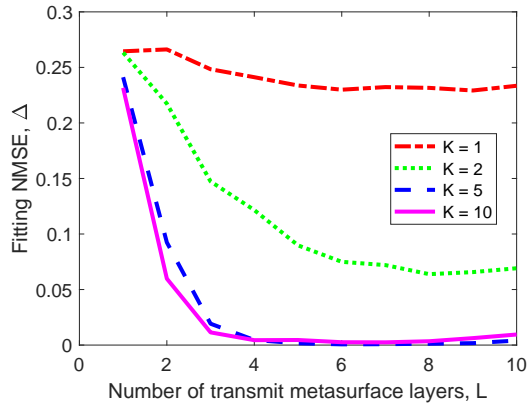}
\caption{The NMSE between the actual channel matrix and the target diagonal one versus the number of TX metasurface layers, where we have $S = 4$, $M = N = 100$, and $r_{e,t} = t_{e,r} = \lambda /2$.}
\label{fig3}
\end{figure}

\begin{figure}[!t]
\centering
\includegraphics[width=7.5cm]{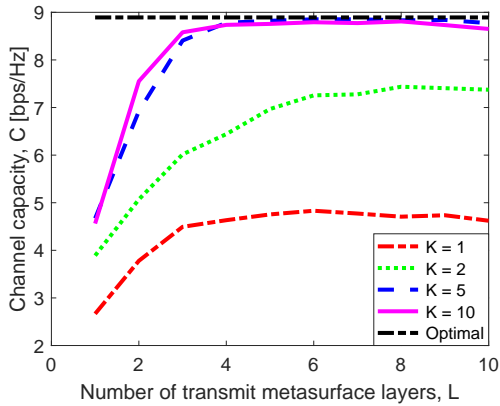}
\caption{The channel capacity versus the number of TX metasurface layers, where we have $S = 4$, $M = N = 100$, and $r_{e,t} = t_{e,r} = \lambda /2$.}
\label{fig4}
\end{figure}

Fig. \ref{fig3} first evaluates the NMSE between the actual channel matrix and the target diagonal matrix for different numbers of metasurface layers, where we consider $S = 4$, $M = N = 100$, and $r_{e,t} = t_{e,r} = \lambda /2$. As such, $A_t$ and $A_r$ are determined accordingly. Observe from Fig. \ref{fig3} that the channel fitting NMSE gradually decreases as the number of metasurface layers increases and eventually bottoms out, when $L \geq 5$ or $K \geq 5$, thanks to the powerful inference capability of the multi-layer architecture advocated. Furthermore, Fig. \ref{fig4} shows the corresponding channel capacity, where the channel capacity of the HMIMO system having a full-precision precoder and combiner is also plotted. It reveals that the SIM-aided HMIMO channel capacity gradually saturates as the number of metasurface layers increases. The SIM having an adequate number of metasurface layers might approach the capacity upper bound characterized by the full-precision precoding and combining. However, due to the fixed thickness of the SIM considered, i.e., $D_{t}$ and $D_{r}$, excessively dense metasurfaces may lead to a performance penalty, when the number of metasurface layers exceeds a certain threshold. For example, the SIM-aided HMIMO using $L = K = 10$ metasurface layers suffers both from some fitting performance erosion as well as from a capacity reduction compared to $L = K = 5$. \emph{In a nutshell, both the channel fitting NMSE and the channel capacity approach their optimal values when using $L = 7$ metasurface layers, and further increasing the number of metasurface layers does not help to improve the fitting NMSE and channel capacity.}


\begin{figure}[!t]
\centering
\includegraphics[width=7.5cm]{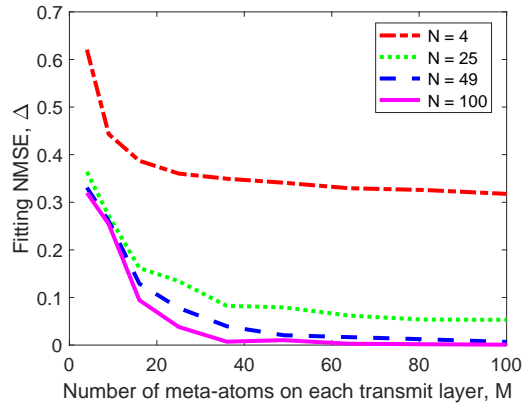}
\caption{The NMSE between the actual channel matrix and the target diagonal one versus the number of meta-atoms on each TX metasurface layer, where we have $S = 4$, $L = K = 7$, and $r_{e,t} = t_{e,r} = \lambda /2$.}
\label{fig5}
\end{figure}
\begin{figure}[!t]
\centering
\includegraphics[width=7.5cm]{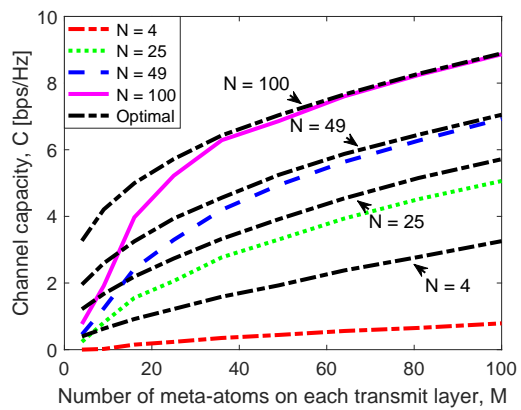}
\caption{The channel capacity versus the number of meta-atoms on each TX metasurface layer, where we have $S = 4$, $L = K = 7$, and $r_{e,t} = t_{e,r} = \lambda /2$.}
\label{fig6}
\end{figure}
In Fig. \ref{fig5}, we portray the channel fitting NMSE versus the number of meta-atoms, where the number of metasurface layers is set to $L = K = 7$, with all other system parameters remaining unchanged. Observe from Fig. \ref{fig5} that the fitting NMSE decreases monotonically as the number of meta-atoms on each transmit or RX metasurface layer increases. As a benefit, the system becomes capable of establishing a perfectly diagonal end-to-end channel matrix for an infinite number of meta-atoms, i.e., $\Delta \rightarrow 0$ for $ M \rightarrow \infty $ or $ N \rightarrow \infty $. Furthermore, Fig. \ref{fig6} plots the channel capacity versus the number of meta-atoms on each TX metasurface layer. Note that as a benefit of the substantial \emph{selection gain} discussed in \emph{Proposition 1}, one can always select the best $S$ subchannels for conveying information \cite{Book_2005_Tse_Fundamentals}. The channel capacity is improved as the number of meta-atoms increases, albeit the number of data streams is fixed. For example, the SIM-aided HMIMO system behaves competitively with its counterpart having full-precision precoding and combiner as $M,\, N \rightarrow 100$. \emph{More specifically, for an adequate number of meta-atoms having tolerable fitting errors, say $N \geq 25$, Fig. \ref{fig6} confirms our Proposition 2 that the channel capacity would increase by about $4$ bps/Hz when doubling the number of meta-atoms in both the TX-SIM and RX-SIM.}

\begin{figure}[!t]
\centering
\includegraphics[width=7.5cm]{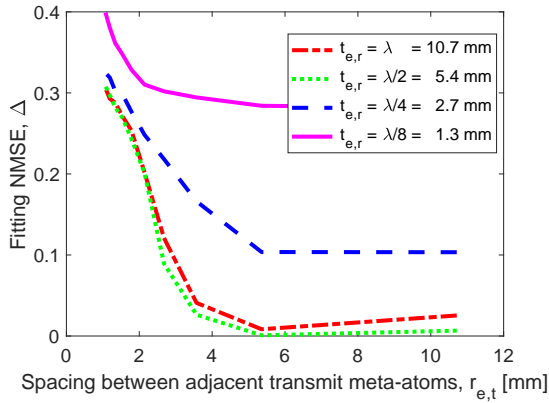}
\caption{The NMSE between the actual channel matrix and the target diagonal one versus the spacing between adjacent meta-atoms on each TX metasurface layer, where we have $S = 4$, $L = K = 7$, and $M = N = 100$.}
\label{fig7}
\end{figure}
\begin{figure}[!t]
\centering
\includegraphics[width=7.5cm]{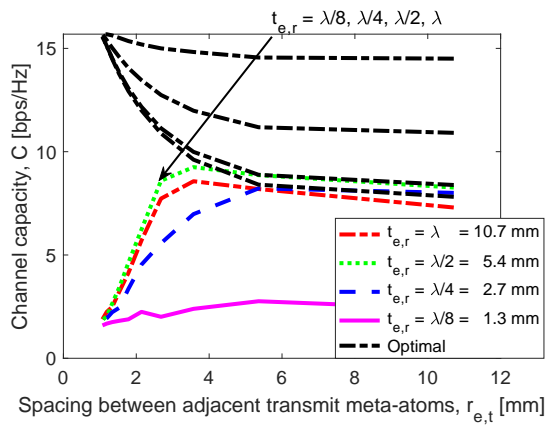}
\caption{The channel capacity versus the spacing between adjacent meta-atoms on each TX metasurface layer, where we have $S = 4$, $L = K = 7$, and $M = N = 100$.}
\label{fig8}
\end{figure}
Furthermore, Figs. \ref{fig7} and \ref{fig8} quantify the channel fitting NMSE and the channel capacity, respectively, versus the spacing between adjacent meta-atoms, where we set $L = K = 7$, $M = N = 100$, and increase the element spacing from $\lambda/10 = 1.1$ mm to $\lambda = 10.7$ mm. It is shown in Fig. \ref{fig7} that the channel fitting NMSE achieves its minimum at $r_{e,t} = t_{e,r} = \lambda /2$. \emph{Both a larger and a smaller element spacing would give rise to the similarity between the transmission coefficients as well as lead to undesired channel correlations, thus resulting in a poor channel fitting NMSE. Observe from Fig. \ref{fig8} that RX-SIM having element spacing of about $t_{e,r} = \lambda/2$ attains the maximal capacity as well as the best fit with the full-precision counterpart.} Given the half-wavelength element spacing between adjacent meta-atoms on each TX metasurface, i.e., $r_{e,t} = \lambda/2$, both the setups of $t_{e,r} = \lambda$ and $t_{e,r} = \lambda/4$ suffer from a capacity loss of about $1$ bps/Hz compared with that adopting $t_{e,r} = \lambda/2$. The inferior fitting NMSE by taking a small element spacing also widens the performance gap between the SIM-assisted HMIMO and its full-precision counterpart. In a nutshell, a better channel fitting NMSE means that parallel subchannels are perfectly formed and suffer from less interference, thus achieving an improved channel capacity for our SIM-aided HMIMO system.

\begin{figure}[!t]
\centering
\includegraphics[width=7.5cm]{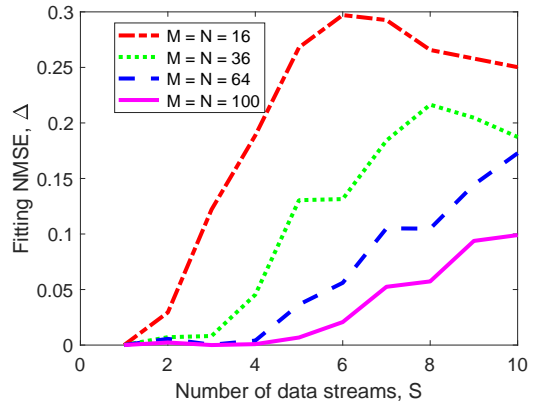}
\caption{The NMSE between the actual channel matrix and the target diagonal one versus the number of data streams, where we have $L = K = 7$, and $r_{e,t} = t_{e,r} = \lambda /2$.}
\label{fig9}
\end{figure}
\begin{figure}[!t]
\centering
\includegraphics[width=7.5cm]{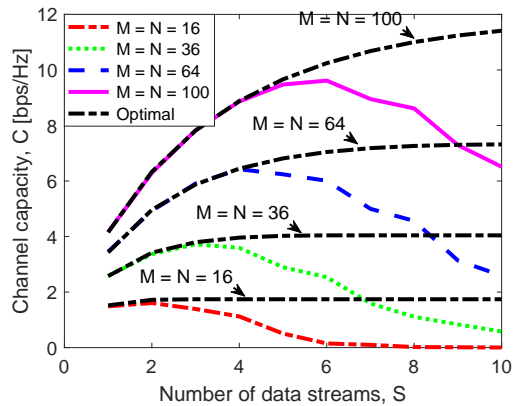}
\caption{The channel capacity versus the number of data streams, where we have $L = K = 7$, and $r_{e,t} = t_{e,r} = \lambda /2$.}
\label{fig10}
\end{figure}
Next, Figs. \ref{fig9} and \ref{fig10} examine the channel fitting NMSE and channel capacity, respectively, versus the number of data streams, where we consider $L = K = 7$, and $r_{e,t} = t_{e,r} = \lambda /2$. It can be seen from Fig. \ref{fig9} that we have $\Delta = 0$ for $S = 1$ under all the setups, and the fitting performance gradually degrades as the number of data streams increases due to the larger dimension of the channel matrix. By utilizing a pair of TX-SIM and RX-SIM having small metasurface profiles, such as $M = N = 16$, we achieve a fitting NMSE of $\Delta = 0.2$ for $S = 4$, which is reduced to $\Delta < 0.001$ upon increasing the number of meta-atoms on each metasurface layer to $M = N = 100$. Furthermore, in sharp contrast to the full-precision counterpart, the SIM-aided HMIMO channel capacity mainly relies on two factors: the number of data streams and the channel fitting NMSE. On one hand, the increasing number of data streams may offer a proportional \emph{multiplexing gain} \cite{Book_2005_Tse_Fundamentals}. On the other hand, it becomes more challenging to acquire a low channel fitting NMSE for a growing number of data streams, thus leading to severe interference among different subchannels. \emph{Due to this fundamental tradeoff between the multiplexing gain and the channel fitting NMSE, the channel capacity achieves its maximum for a certain number of data streams, e.g., $S = 2,\, 3,\, 4,\, 6$ for the four setups considered in Fig. \ref{fig10}.} In addition, note that the ideal channel capacity approaches saturation as the number of data streams increases, which is consistent with our \emph{Proposition 1}.

\subsection{Validation of the Proposed Algorithm}
\begin{figure}[t!]
	\centering
	\subfloat[\label{fig11a}$L = K = 1$;]{
		\includegraphics[width=4.3cm]{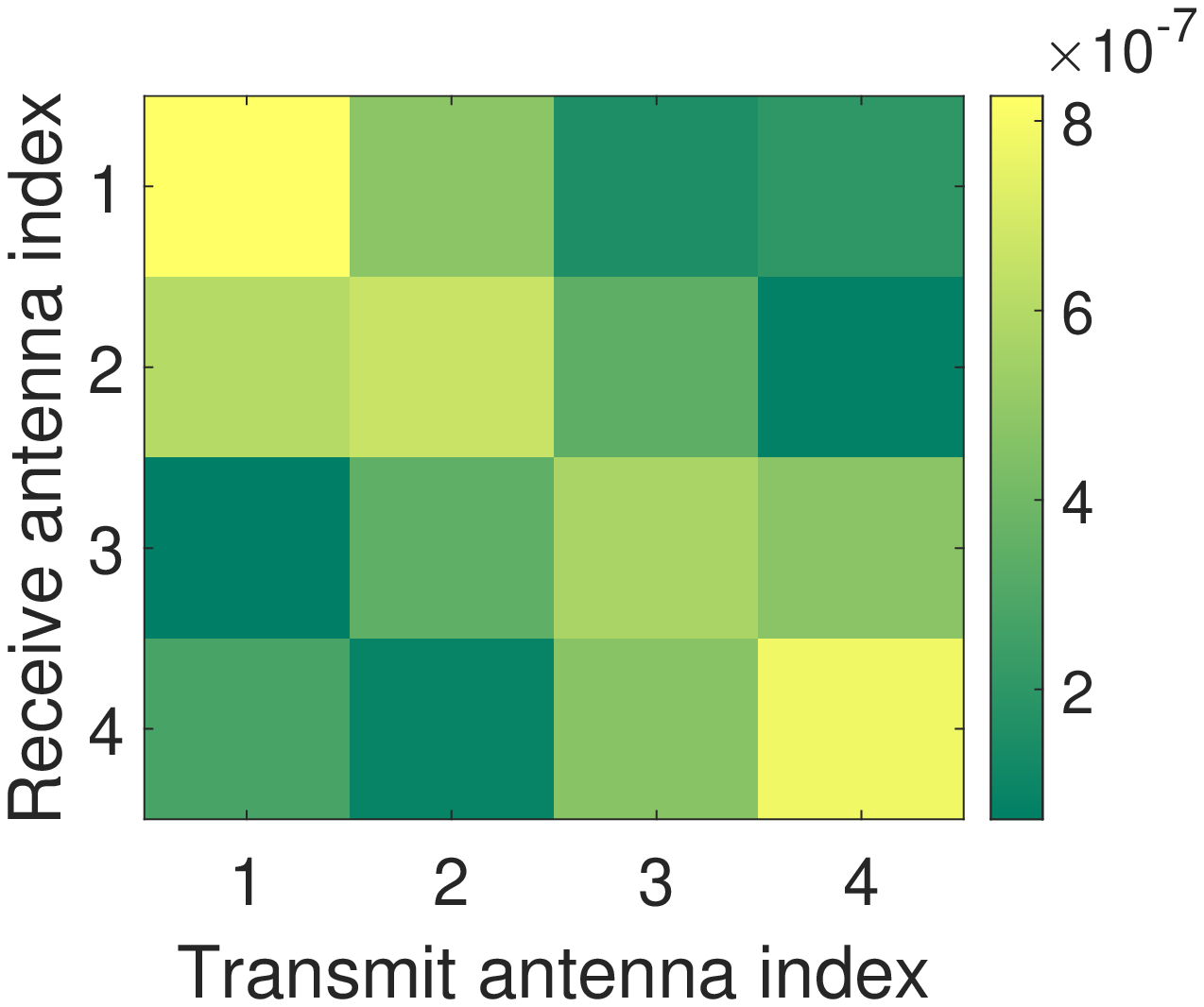}}
	\subfloat[\label{fig11b}$L = K = 2$;]{
		\includegraphics[width=4.3cm]{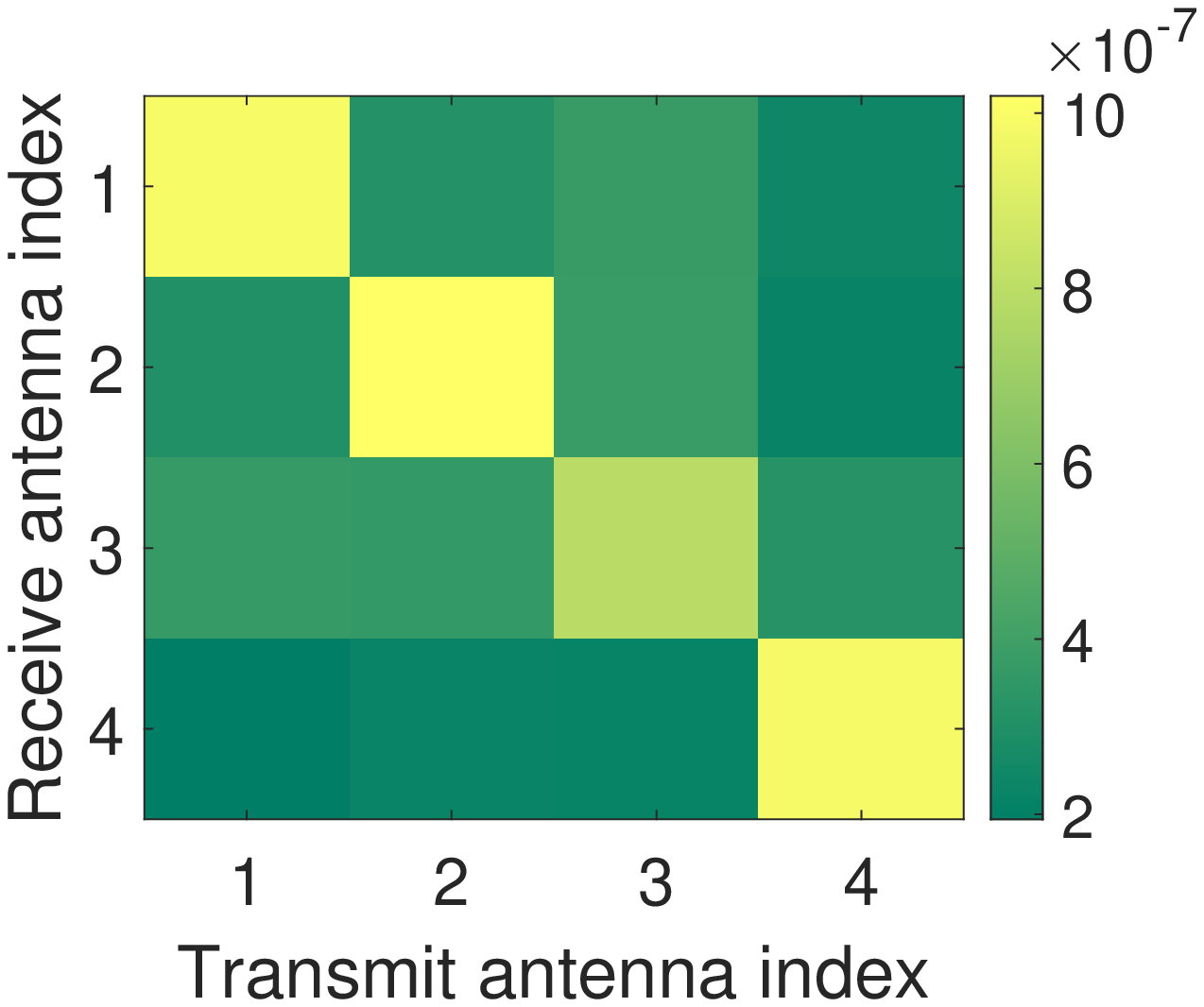}}
	\\
	\subfloat[\label{fig11c}$L = K = 3$;]{
		\includegraphics[width=4.3cm]{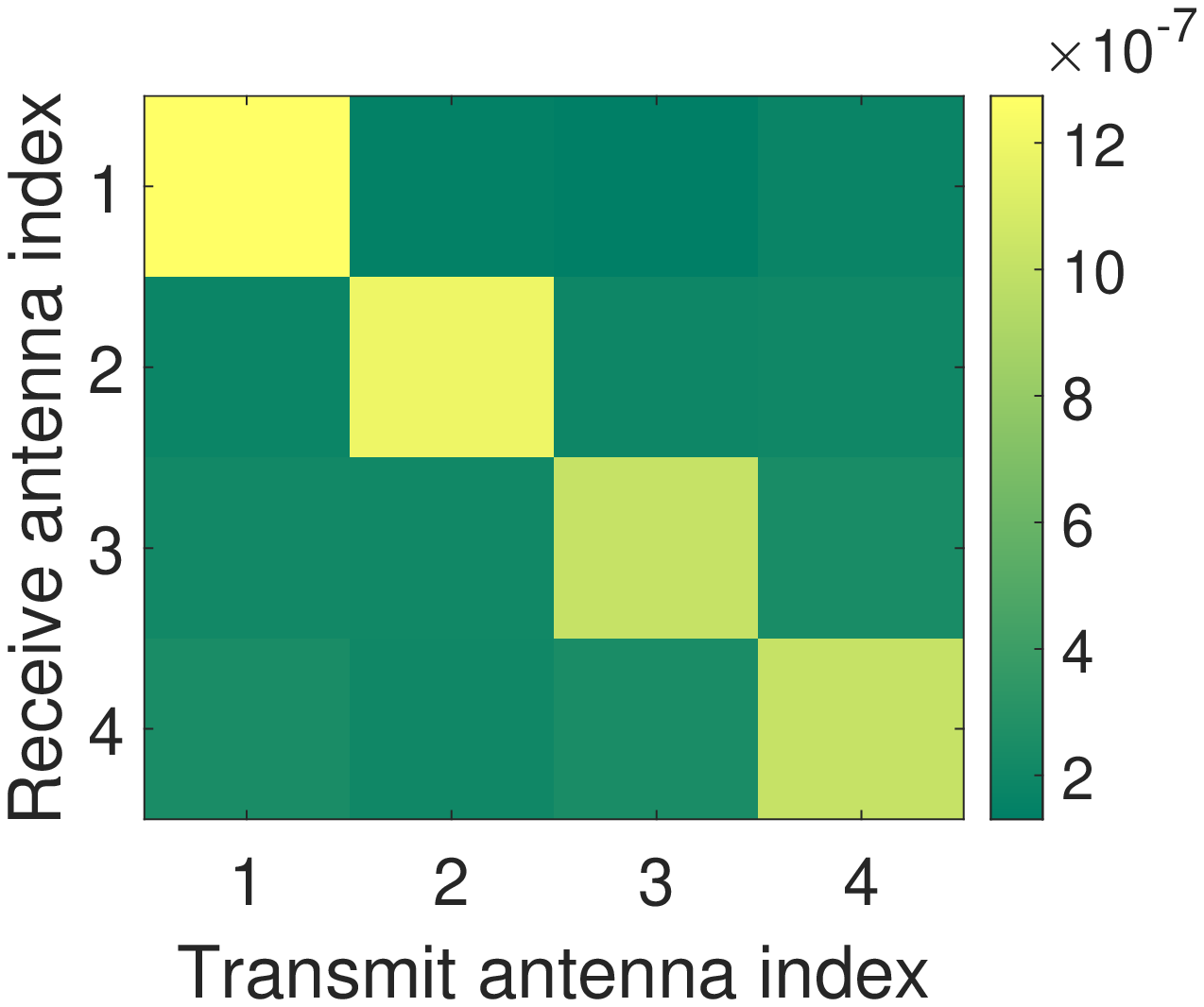}}
	\subfloat[\label{fig11d}$L = K = 4$;]{
		\includegraphics[width=4.3cm]{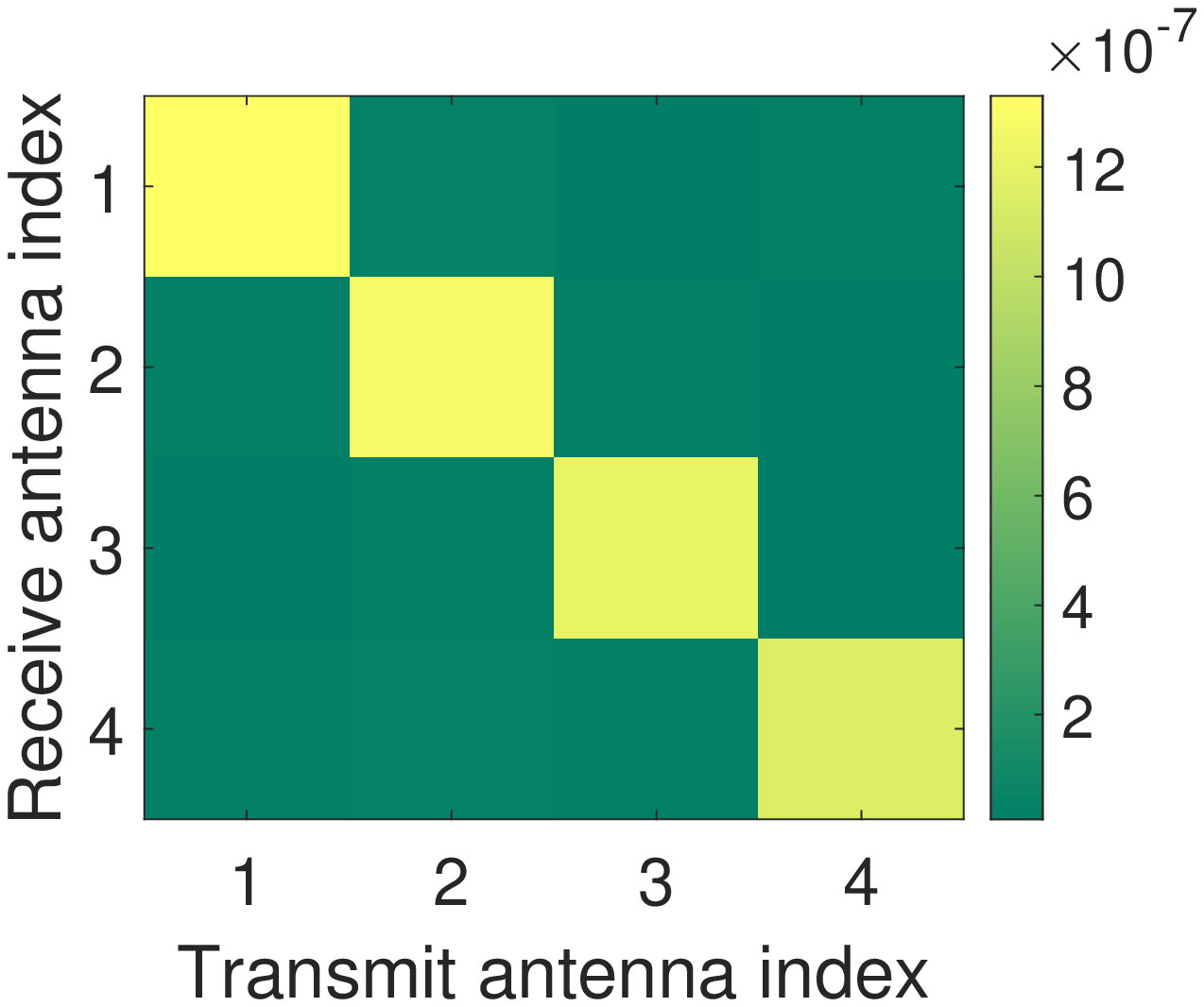}}
	\caption{The visualization of the end-to-end spatial channel matrix $\mathbf{H}=\mathbf{Q}\mathbf{G}\mathbf{P}$.}
	\label{fig11} 
\end{figure}
\begin{figure}[t!]
	\centering
	\subfloat[\label{fig12a}$\eta_{0} = 0.1$;]{
		\includegraphics[width=4.3cm]{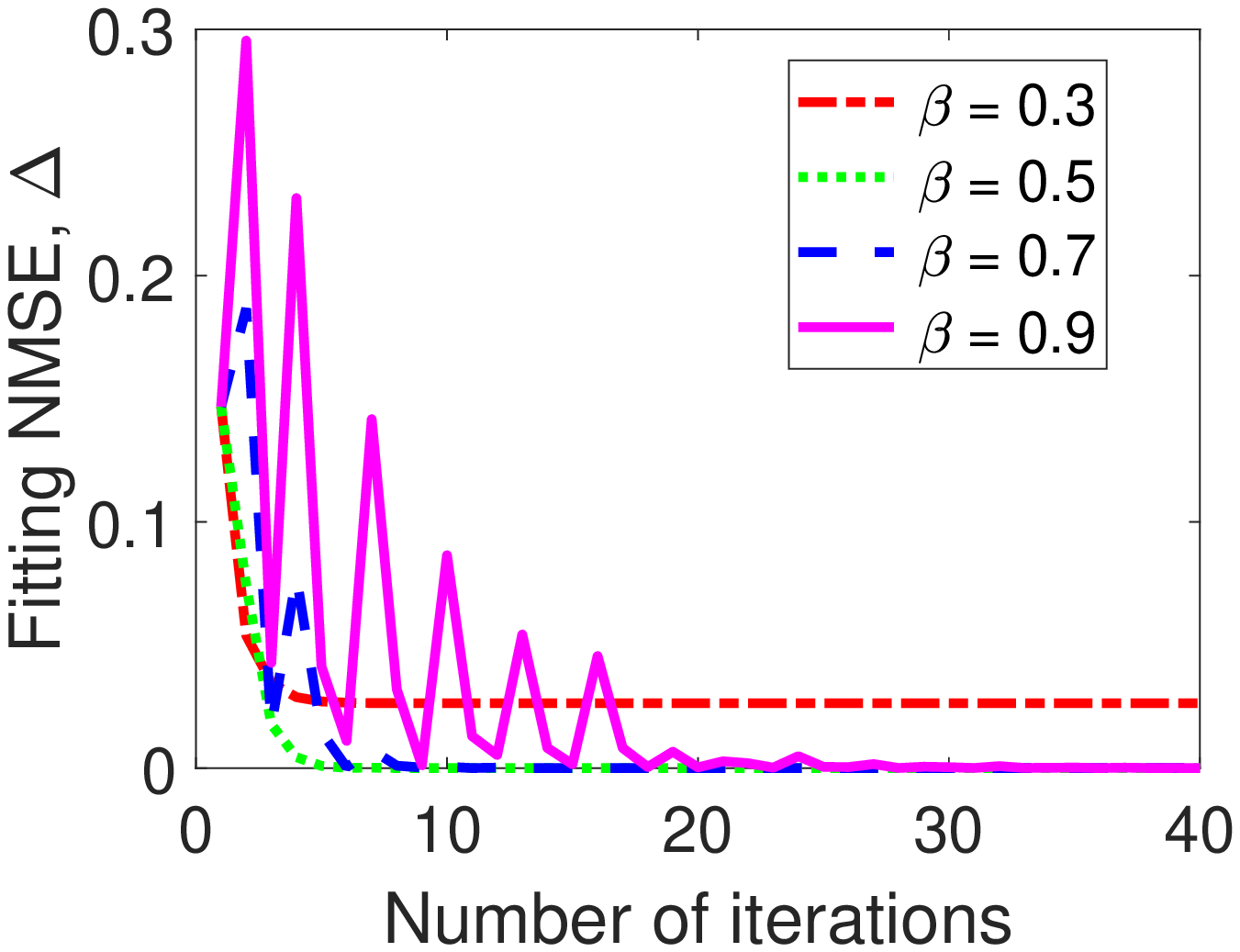}}
	\subfloat[\label{fig12b}$\eta_{0} = 0.2$;]{
		\includegraphics[width=4.3cm]{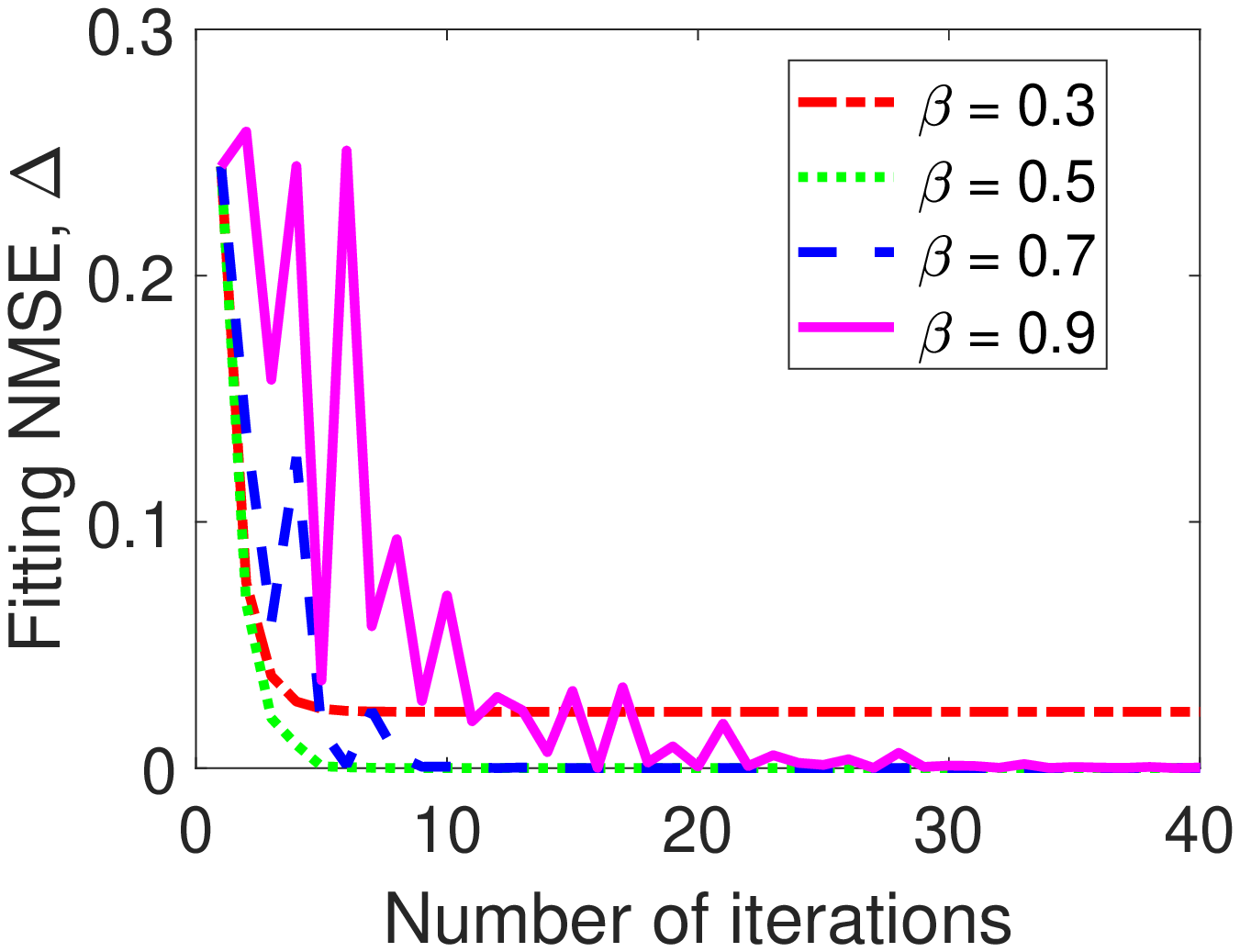}}
	\\
	\subfloat[\label{fig12c}$\eta_{0} = 0.5$;]{
		\includegraphics[width=4.3cm]{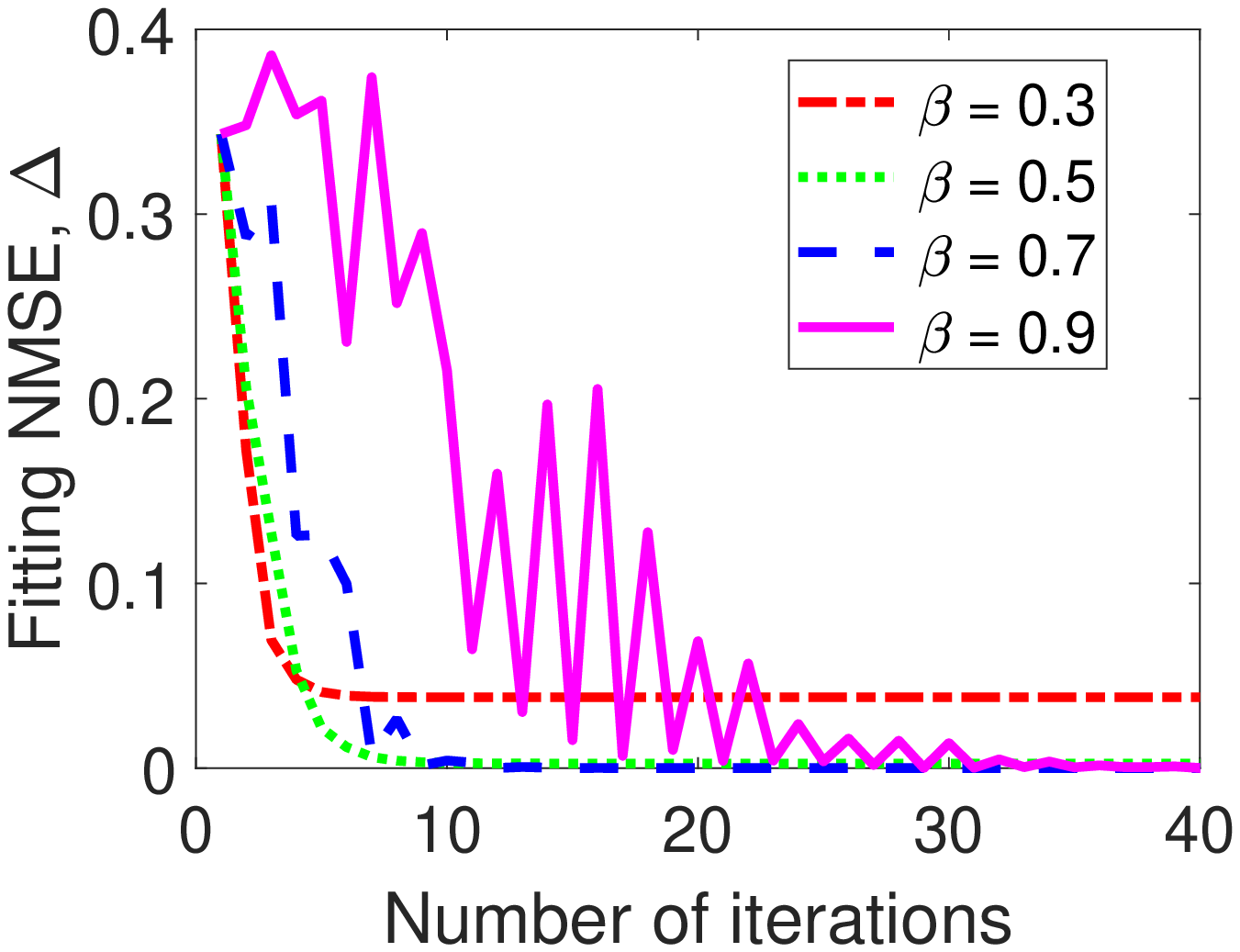}}
	\subfloat[\label{fig12d}$\eta_{0} = 1.0$;]{
		\includegraphics[width=4.3cm]{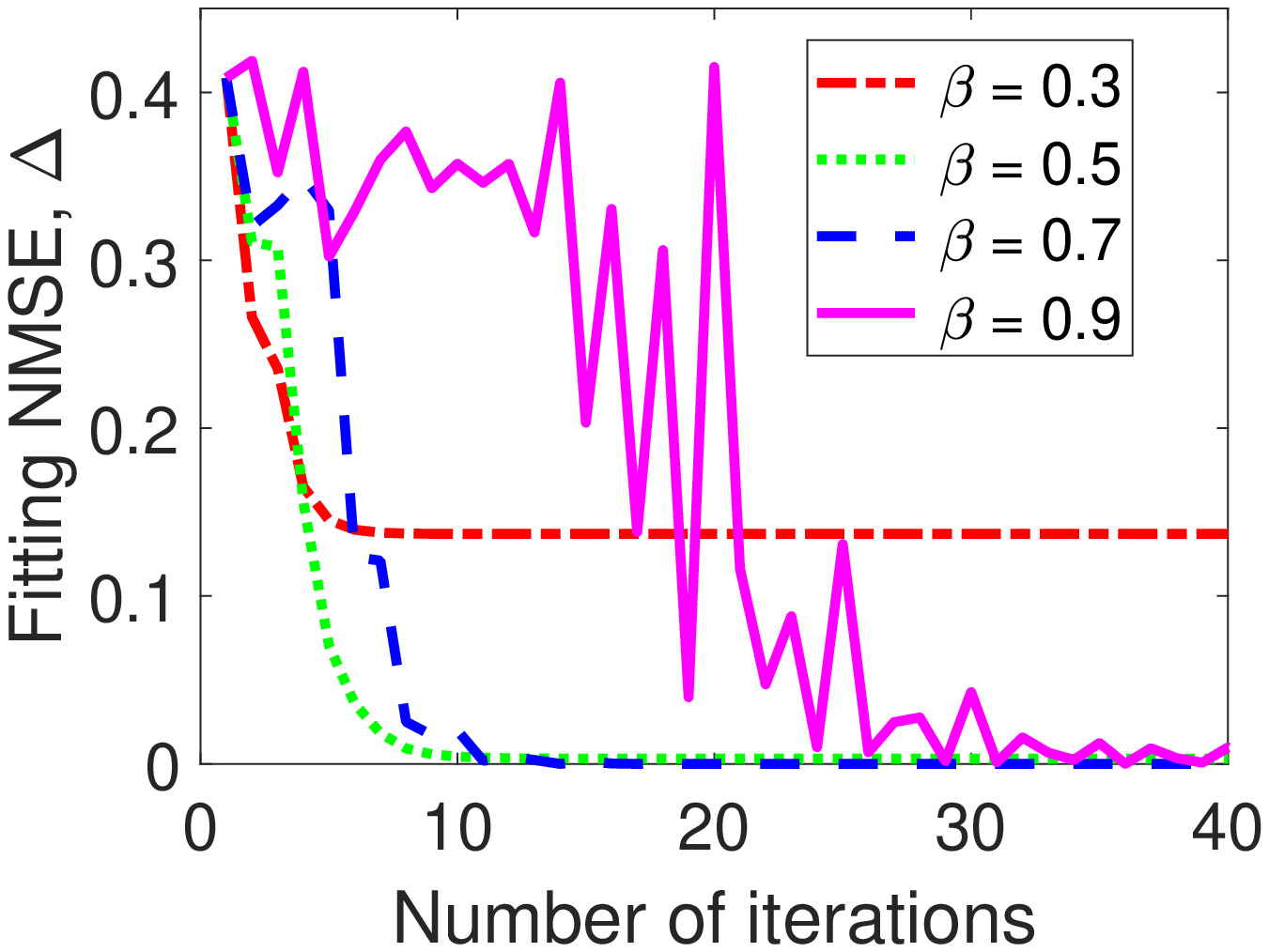}}
	\caption{The convergence curves of the proposed gradient descent algorithm.}
	\label{fig12} 
\end{figure}
For the sake of illustration, Fig. \ref{fig11} visualizes the end-to-end channel matrix $\mathbf{H}=\mathbf{Q}\mathbf{G}\mathbf{P}$ for different numbers of metasurface layers, where we consider $S=4$, $M = N = 100$, and $r_{e,t} = t_{e,r} = \lambda /2$. Observe from Fig. \ref{fig11} that for a small number of metasurface layers, such as $L = K =1$, the TX-SIM and RX-SIM struggle to form a diagonal channel matrix spanning from the source to the destination. Hence, each data stream suffers from the interference imposed by other streams, ultimately resulting in a reduced channel capacity (see Fig. \ref{fig4}). As the number of metasurface layers increases, the TX-SIM and RX-SIM attain a stronger inference capability and thus may form multiple parallel subchannels in the physical space. Fig. \ref{fig11}(d) shows that the TX-SIM and RX-SIM having four metasurface layers respectively succeed in forming an almost perfectly diagonal channel matrix, thus asymptotically achieving the maximal channel capacity.

Next, we examine the convergence performance of the proposed gradient descent algorithm by considering different values of the initial learning rate $\eta_{0}$ and the decay parameter $\beta$. As shown in Fig. \ref{fig12}(a), we begin by analyzing the case of $\eta_{0} = 0.1$. Note that under all the setups, the fitting NMSE eventually decreases, thus facilitating convergence. However, for a small value of the decay parameter, e.g., $\beta = 0.3$, the channel fitting NMSE might converge to a local minimum. By contrast, a larger value of the decay parameter, e.g., $\beta = 0.9$, may result in overshooting effects. As a result, the fitting NMSE fluctuates violently during the initial iteration stage. Furthermore, when we increase the initial learning rate to $\eta_{0} = 0.2,\, 0.5,\, 1.0$, the corresponding results are shown in Figs. \ref{fig12}(b), \ref{fig12}(c), and \ref{fig12}(d), respectively. It is demonstrated that an excessively high initial learning rate may require a long period to achieve convergence. For example, more than $40$ iterations are required for the setup of $\eta_{0} = 1.0$ and $\beta = 0.9$. Nonetheless, in all cases, the fitting NMSE can converge to the desired accuracy after a sufficient number of iterations.
\begin{figure}[!t]
\centering
\includegraphics[width=7.5cm]{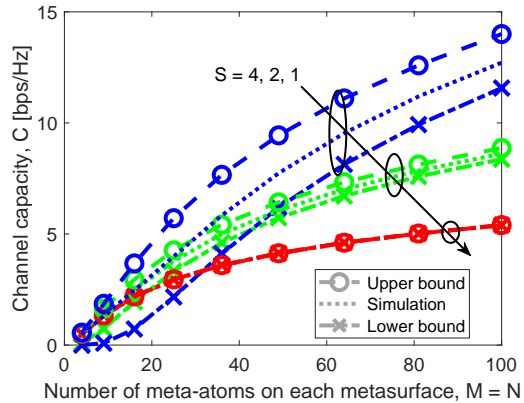}
\caption{Channel capacity comparison of the simulation and analytical results based on \eqref{eq33}.}
\label{fig13}
\end{figure}

\subsection{Performance Evaluation and Comparison to Existing Transmission Technologies}
Furthermore, Fig. \ref{fig13} verifies the accuracy of our theoretical analysis of the HMIMO channel capacity, where we consider $r_{e,t} = t_{e,r} = \lambda /4$ and increase the number of data streams from $S = 1$ to $S = 4$. Observe from Fig. \ref{fig13} that the channel capacity increases with the number of meta-atoms on each metasurface as well as that of data streams, which is due to the substantial \emph{selection gain} and \emph{multiplexing gain}, respectively \cite{Book_2005_Tse_Fundamentals}. Specifically, when considering $S = 4$ data streams, the $8$ bps/Hz capacity increase is observed by quadrupling the number of meta-atoms on each metasurface from $M = N = 25$ to $M = N = 100$, which is consistent with our previous analysis in \emph{Proposition 2}. Furthermore, as expected, the actual channel capacity of our HMIMO communication system consistently lies between the upper and lower bounds derived. Specifically, both the upper and lower bounds are tight for a single data stream, i.e., $S = 1$. As the number of data streams increases, there is a widening gap between the analytical and simulation results due to the imperfect scaling operation in \eqref{eq33}. Deriving the accurate capacity calls for future research.

\begin{figure}[!t]
\centering
\includegraphics[width=7.5cm]{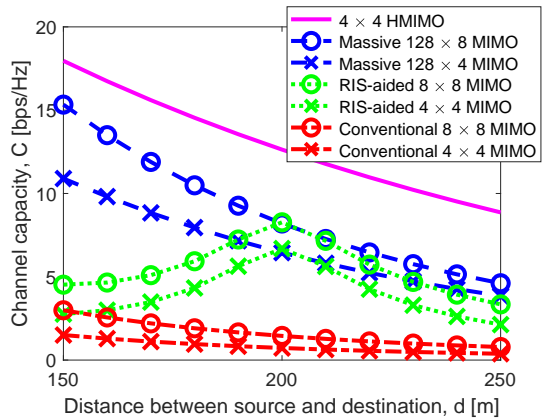}
\caption{Channel capacity comparison of our SIM-aided HMIMO ($S = 4$, $L = K = 7$, $M = N = 100$, $r_{e,t} = t_{e,r} = \lambda /2$) and other MIMO transmission schemes.}
\label{fig14}
\end{figure}
\begin{figure}[!t]
\centering
\includegraphics[width=7.5cm]{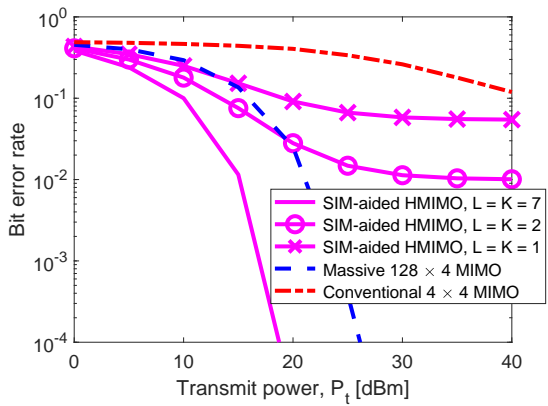}
\caption{Bit error rate comparison of our SIM-aided HMIMO ($S = 4$, $M = N = 100$, $r_{e,t} = t_{e,r} = \lambda /2$) and other MIMO transmission schemes, where the transmission rate is 4 bpcu.}
\label{fig14_0}
\end{figure}
In Fig. \ref{fig14}, we compare the channel capacity of our SIM-assisted HMIMO system to that of the massive MIMO scheme as well as to its RIS-aided MIMO counterpart. The detailed MIMO setups are shown in Fig. \ref{fig14}. Specifically, we adopt a pair of TX-SIM and RX-SIM ($S = 4$, $L = K = 7$, $M = N = 100$, $r_{e,t} = t_{e,r} = \lambda /2$) for performing the wave-based precoding and combining, while achieving the spatial gains. As for the RIS-aided MIMO scheme, a RIS having 1,000 elements is deployed at a site having a source-RIS distance of $200$ m and a vertical spacing of $10$ m \emph{w.r.t.} the source-destination link to enhance the channel quality. Thus, we have a RIS-destination distance of $\sqrt{50^{2}+10^{2}} \approx 51$ m. All the channels are assumed to be Rayleigh fading along with the path loss model in \eqref{eq16}. The path loss exponents are adjusted to $b = 2.2$ and $b = 2.7$ for the source-RIS and RIS-destination links, respectively \cite{JSAC_2020_Zhang_Capacity}. Observe from Fig. \ref{fig14} that as a benefit of the substantial spatial gain attained by the TX-SIM and RX-SIM, our HMIMO outperforms both its massive MIMO and RIS-aided counterparts under all the setups considered. Specifically, although RIS achieves significant capacity improvements over the conventional MIMO, it still suffers from a performance gap compared to the HMIMO due to the severe two-hop path loss. Even in the vicinity of RIS, the HMIMO attains a 150\% capacity gain, which may increase to 200\% at the cell edge, e.g., at $d = 250$ m. Additionally, the massive MIMO equipped with a huge number of active elements achieves impressive capacity improvements at the cost of an increasing number of active RF chains, which, however, still has at least a $3$ bps/Hz capacity penalty compared to the HMIMO scheme.

\begin{figure}[!t]
\centering
\includegraphics[width=7.5cm]{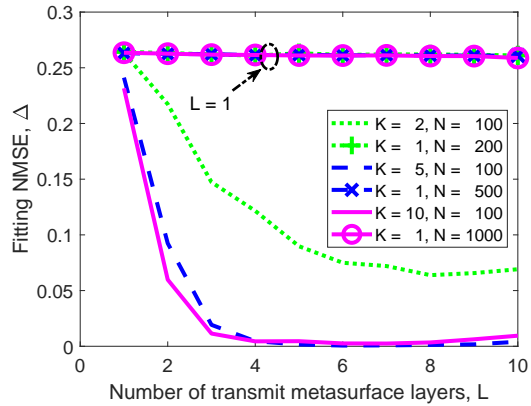}
\caption{The channel fitting NMSE comparison of the multilayer SIM and its single-layer counterpart that has the same total number of meta-atoms.}
\label{fig15}
\end{figure}
\begin{figure}[!t]
\centering
\includegraphics[width=7.5cm]{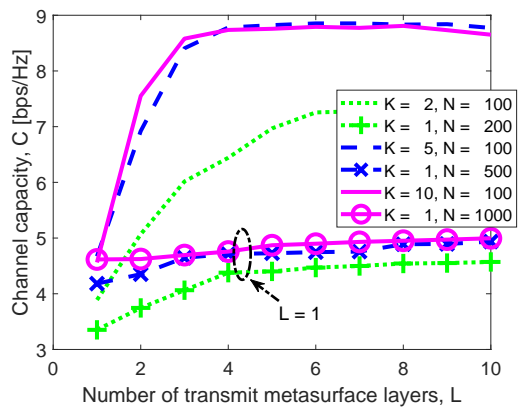}
\caption{The channel capacity comparison of the multilayer SIM and its single-layer counterpart that has the same total number of meta-atoms.}
\label{fig16}
\end{figure}
Fig. \ref{fig14_0} compares the error performance of our SIM-assisted HMIMO system to conventional MIMO schemes. Specifically, we consider four data streams, each transmitting a BPSK symbol, which corresponds to a transmission rate of 4 bits per channel use (bpcu). The distance between the transmitter and receiver is set to 200 m, and the number of metasurface layers is set to $L = K = 1,\ 2,\ 7$, respectively. As shown in Fig. \ref{fig14_0}, deploying a sufficient number of metasurface layers, such as $L = K =7$ in both the TX-SIM and RX-SIM, effectively mitigates the inter-stream interference, thanks to carrying out the precoding and combining in the wave domain. Furthermore, as a benefit of the spatial gain attained by the large transceiver surface aperture, the SIM-assisted HMIMO achieves a lower bit error rate (BER) than its large-scale MIMO counterpart. Observe from Fig. \ref{fig14_0} that SIM-aided HMIMO has an $8$ dB performance gain compared to massive MIMO in this setup. However, when reducing the number of metasurface layers to $L = K = 2$, the TX-SIM and RX-SIM modules failed to perfectly suppress the interference amongst the data streams. Consequently, the SIM-assisted HMIMO scheme suffers from performance erosion, which results in a residual BER as the transmit power $P_t$ increases. Note that we directly use the SIM phase shifts optimized in Section \ref{sec32} to evaluate the BER performance and that directly optimizing the SIM for minimizing the BER may result in further performance improvements.

Finally, Fig. \ref{fig15} compares the channel fitting NMSE of the multilayer SIM to its single-layer counterpart. Specifically, the total number of meta-atoms in the TX-SIM and RX-SIM are rearranged into a single-layer metasurface, respectively. In order to maintain the same transceiver surface area, the meta-atom spacing is set to $r_{e,t} = 5\lambda /\left \lceil \sqrt{ML} \right \rceil$ and $t_{e,r} =5\lambda /\left \lceil \sqrt{NK} \right \rceil$. All other simulation parameters are consistent with those in Fig. \ref{fig3}. Observe from Fig. \ref{fig15} that the single-layer SIM fails to accurately fit the expected end-to-end channel, even with an adequate number of meta-atoms. By contrast, the multilayer SIM structure achieves a superior channel fitting NMSE as the number of metasurface layers increases. Furthermore, Fig. \ref{fig16} shows the channel capacity of these two transmission schemes. It is evident that the single-layer SIM provides only marginal capacity improvements as the number of meta-atoms increases, which is primarily due to the inability of the single-layer TX-SIM and RX-SIM to effectively suppress the inter-stream interference. Moreover, deploying a large number of meta-atoms in a limited space also leads to channel correlation. Observe from Fig. \ref{fig16} that the proposed SIM-aided HMIMO system associated with $L = K = 10$ and $M = N = 100$ achieves almost twice the capacity improvement compared to its single-layer counterpart for the same total number of meta-atoms. In a nutshell, both the channel fitting NMSE and the corresponding channel capacity of the single-layer SIM suffer from significant performance penalties compared to its multi-layer SIM counterpart. Nevertheless, finding the optimal SIM design under a given total number of meta-atoms remains an open research question that requires further investigation.

\section{Conclusions}\label{sec7}
In this paper, we proposed a novel SIM-based HMIMO communication paradigm, which attains substantial spatial gains while performing the precoding and combining functionalities directly in the native EM regime at the speed of light. We first formulated a channel fitting problem to approximate the MIMO-capacity-optimal diagonal channel matrix by optimizing the phase shifts of both the TX-SIM and RX-SIM. Then, we proposed an efficient gradient descent algorithm for iteratively solving that non-trivial fitting problem. Additionally, we derived a numerical approximation method for characterizing the HMIMO channel capacity and derived some fundamental capacity scaling laws. Finally, extensive simulations were provided for validating the benefits of the proposed SIM-based HMIMO system, demonstrating that substantial capacity improvements were attained upon increasing the number of the SIMs' meta-atoms.

In conclusion, our pivotal findings are as follows. Firstly, our experimental insights have shed light on the optimal SIM design. Specifically, we found that a $7$-layer SIM having half-wavelength element spacing achieves an excellent channel fitting performance approaching the MIMO channel capacity. Secondly, both our theoretical analysis and simulation results have shown a quadratic channel gain when doubling the number of meta-atoms. Additionally, we have verified the performance advantages of the proposed HMIMO scheme over the existing benchmark schemes. Notably, a 150\% capacity gain was attained over its conventional massive MIMO and RIS-assisted counterparts. As such, the multilayer SIM structure is capable of carrying out signal processing in the wave domain, which might lead to disruptive implementation-oriented advances. Moreover, an active SIM may be created by integrating small power amplifiers within some of the meta-atoms \cite{NE_2022_Liu_A}. Upon adjusting the drive level of these power amplifiers, a non-linear module can be produced for further enhancing the inference capability of the SIM. Nonetheless, accurately evaluating the achievable performance gain of the active SIM requires further investigation.

\bibliographystyle{ieeetr}
\bibliography{ref}
\end{document}